\documentclass[reqno,11pt]{article}
\usepackage{psfig, amsmath, amstexnb, amsthm}
\input control.sty

\begin{document}


\begin{titlepage}

\title{Control of Dynamic Hopf Bifurcations}

\author{
N. Berglund\\
{\it Weierstra\ss-Institut f\"ur Angewandte Analysis und Stochastik} \\
{\it Mohrenstra\ss e 39, D-10117 Berlin, Germany} \\
}

\date{March 30, 1999}

\maketitle

\begin{abstract}
The slow passage through a Hopf bifurcation leads to the delayed
appearance of large amplitude oscillations. We construct a smooth scalar
feedback control which suppresses the delay and causes the system to follow
a stable equilibrium branch. This feature can be used to detect in time the
loss of stability of an ageing device. As a by-product, we obtain results
on the slow passage through a  bifurcation with double zero eigenvalue,
described by a singularly perturbed cubic Li\'enard equation.
\end{abstract}

\vspace{\fill}

\noindent
\emph{1991 Mathematics Subject Classification.} 34E15, 58F14, 93D15.

\noindent
\emph{Key words and phrases.} Hopf bifurcation, nonlinear control theory,
singular perturbations, dynamic bifurcations, codimension four unfolding,
Li\'enard equation.

\thispagestyle{empty}

\end{titlepage}


\section{Introduction}
\label{sec_in}

The main motivation for this paper is the problem of ageing of dynamical
systems. Assume that the behaviour of a device can be described by an 
$n$--dimensional ordinary differential equation $\dot{x} = f(x)$. The system
is designed in such a way that $\fix{x}$ is an asymptotically stable
equilibrium point, which corresponds to the desired behaviour. Slow changes
of the system's characteristics due to the ageing process can be modelled by
a slowly time--dependent equation
\begin{equation}
\label{in1}
\dtot{x}{t} = f(x,\eps t), \qquad 0<\eps\ll 1,
\end{equation}
where $f(x,0)$ describes the dynamics of the brand new device.

Equation \eqref{in1} being a nonautonomous differential equation which is
difficult to solve, one is tempted to consider instead the 1--parameter
family of autonomous systems
\begin{equation}
\label{in2}
\dtot{x}{t} = f(x,\lambda), \qquad \lambda = \text{constant.}
\end{equation}
One hopes that if the ``quasistatic approximation'' \eqref{in2} has a family
of attractors depending smoothly on $\lambda$, then solutions of \eqref{in1}
should be close, at any given time $t$, to the attractor of \eqref{in2} with
$\lambda=\eps t$. 

This is at least partially justified by the following result
\cite{PR,VBK,Fe}: if $\fix{x}(\lambda)$ is a family of asymptotically
stable equilibria of \eqref{in2}, then any solution of \eqref{in1} starting
in a sufficiently small \nbh\ of $\fix{x}(0)$ will, after a short
transient, track the curve $\fix{x}(\eps t)$ at a distance of order $\eps$.
For the ageing device, this implies that we need not worry as long as the
``nominal'' equilibrium $\fix{x}(\lambda)$ remains asymptotically stable. 

This naturally raises the question of what happens if the equilibrium
$\fix{x}(\lambda)$ undergoes a bifurcation at $\lambda=\lambda_0$. Such
problems are usually referred to as \defwd{dynamic bifurcations}
\cite{Benoit}. Bifurcations with a single zero eigenvalue have been studied
in some detail. It turns out that saddle--node bifurcations will result in
a sudden jump of the solution \cite{Hab}, which may have catastrophic
consequences for the device. Transcritical and pitchfork bifurcations
generically result in a smoother behaviour, where the trajectory follows one
of the stable equilibria created in the bifurcation \cite{LS}. This feature
might be used to detect the loss of stability of the nominal equilibrium, in
order to switch off the device before any harm is done. 

The case of a Hopf bifurcation has been analysed more recently
\cite{Sh,Ne,Ne2}. The surprising phenomenon is that instead of directly
tracking the limit cycle created in the bifurcation, the trajectory remains
close to the unstable equilibrium for some time, before jumping to the
periodic orbit. Instead of oscillations with a continuously increasing
amplitude, one thus observes the sudden appearance, after some delay, of
large amplitude oscillations (\figref{fig_pro}a). The bifurcation delay is
stable with respect to smooth deterministic perturbations. For the ageing
device, this phenomenon implies that the loss of stability cannot be
detected soon enough to avoid catastrophic oscillations. 

To avoid such problems, one may try to control the system. A
simple affine control of the quasistatic system \eqref{in2} would be 
\begin{equation}
\label{in3}
\dtot{x}{t} = f(x,\lambda) + b\, u,
\end{equation}
where $b$ is a given vector in $\R^n$ and $u$ is a scalar function. Two
cases have been studied:
\begin{enum}
\item	\defwd{Open loop control:} $u(t)$ is a function of time taking
values in some compact interval $U$. The analysis of \eqref{in3} for all
possible functions of this type leads to the notion of control sets, which
tend to form around invariant sets. The $\lambda$--dependence of control
sets near bifurcation points has been studied in \cite{CK,CHK,HS}. 

\item	\defwd{Feedback control:} $u(x(t))$ is a function of the state of the
system. The bifurcation is ``stabilized'' by choosing the function $u(x)$ in
such a way that the nominal equilibrium of \eqref{in3} is stable when
$\lambda=\lambda_0$. For topological reasons, this makes the bifurcation
supercritical, and should avoid exploding trajectories at least for
$\lambda$ slightly larger than $\lambda_0$ \cite{Ab,Ae,MS}. 
\end{enum}

The sometimes surprising behaviour of dynamic bifurcations shows that we
should consider the slowly time-dependent version of \eqref{in3} as well.
Here we will limit ourselves to feedback controlled systems of the form
\begin{equation}
\label{in4}
\dtot{x}{t} = f(x,\eps t) + b\, u(x,\eps t)
\end{equation}
in the specific case where $f$ undergoes a Hopf bifurcation. Since we wish
to analyse \eqref{in4} on the time scale $\eps^{-1}$, we introduce the
\defwd{slow time} $\tau = \eps t$ and rewrite \eqref{in4} as the singular
perturbation problem
\begin{equation}
\label{in5}
\eps\dtot{x}{\tau} = f(x,\tau) + b\, u(x,\tau).
\end{equation}
Our aim is to design a feedback control $u$ in such a way that the
bifurcation delay, and hence the sudden appearance of large amplitude
oscillations, are suppressed. 

A feedback affecting only nonlinear terms in $x-\fix{x}$ will have no effect
on the delay. We thus have to modify the linearization $A$ of $f$ at the
bifurcation point. Shifting the real part of the eigenvalues of $A$ will
merely postpone the problem to some later time. The only solution is thus to
shift the imaginary part of the eigenvalues in order to produce a double
zero eigenvalue. 

Our strategy will thus be the following. First we design a feedback in such
a way that when the parameter $\lambda$ is varied through the bifurcation
value $\lambda_0$, new stable equilibrium branches are created at a
distance of order $(\lambda-\lambda_0 )^{1/2}$ of the nominal branch
$\fix{x}(\lambda)$. The resulting vector field turns out to be a
codimension four unfolding studied in \cite{KKR}.  We then show that the
solutions of the corresponding system \eqref{in5} actually track one of
these branches, a feature which can be used to detect the bifurcation point.

To analyse the time-dependent system, we extend certain methods of
\cite{B1} (see \cite{B2} for a summary and \cite{BK} for applications).
However, this paper is written in a largely self-contained way. It is
organized as follows: the main result is stated in Section \ref{sec_pro};
in Section \ref{sec_ucp} we recall Neishtadt's result on bifurcation delay
\cite{Ne}, which will guide us in the construction of the feedback in
Section \ref{sec_bif}.  Section \ref{sec_sdp} is devoted to the proof of
the main result for the singularly perturbed equation \eqref{in5}. 
Finally, in Section \ref{sec_qpr}, we give a few remarks on what happens
when the control is slightly imperfect, and the requirements of the theorem
are no longer met.

\vspace{\fill}

\noindent
{\it Acknowledgments:} It is a pleasure to thank Klaus Schneider for
welcoming me at the Weierstra\ss\ Institute and for inspiring discussions
on this interesting topic. I thank Dima Turaev for pointing out useful
references on high codimension bifurcations. This work is supported by the
Nonlinear Control Network of the European Community, Grant ERB
FMRXCT--970137.


\newpage
\section{The Problem and Main Result}
\label{sec_pro}

We consider the feedback controlled dynamical system
\begin{equation}
\label{pro1}
\dtot{x}{t} = f(x,\lambda) + b \, u(x,\lambda), 
\qquad x\in\R^n, \; \lambda\in\R
\end{equation}
where:

\begin{itemiz}
\item	The \defwd{uncontrolled vector field} $f(x,\lambda)$ undergoes a
	Poincar\'e--Andronov--Hopf bifurcation at the origin.
		
\item	The vector $b\in \R^n$ is imposed and fixed. It describes the
	direction in which the system can be steered.
	
\item	The \defwd{scalar feedback control} $u(x,\lambda)\in\R$ is a
	function to be determined in such a way that the solution behaves 
	``smoothly'' when $\lambda$ is slowly varied. Its dependence on $x$
	and $\lambda$ should be as simple as possible (\eg polynomial).   
\end{itemiz}

More precisely, we will assume that the uncontrolled vector field satisfies
the following hypotheses:

\begin{myhyp}{(H1)}{Domain and smoothness}
The function $f(x,\lambda): \cD\times I\to \R^n$ is analytic in a \nbh\
$\cD$ of the origin in $\R^n$ and an interval $I$ containing 0.
\end{myhyp}

\begin{myhyp}{(H2)}{Hopf bifurcation}
There exists a curve $\fix{x}(\lambda): I\to\R^n$ with $\fix{x}(0)=0$ and
$f(\fix{x}(\lambda),\lambda)=0$. The Jacobian matrix $A(\lambda) =
\sdpar{f}{x}(\fix{x}(\lambda), \lambda)\in \R^{n\times n}$ admits two
eigenvalues $a(\lambda)\pm\icx\w(\lambda)$, where $a(0)=0$, $a'(0)>0$ and
$\w(0)=\w_0\neq 0$. All other eigenvalues of $A(\lambda)$ have a strictly
negative real part.
\end{myhyp}

These hypotheses imply in particular that there is a set of coordinates  $x
= (y,z)$, with $y\in\R^m$ ($m=n-2$) and $z=(\x,\y)\in\R^2$, such that 
\begin{equation}
\label{pro2}
f(x,0) = 
\begin{pmatrix}
A_- y + g_-(y,z) \\
A_0 z + g_0(y,z)
\end{pmatrix}, \qquad
A_0 = 
\begin{pmatrix}
0 & \w_0 \\ -\w_0 & 0
\end{pmatrix},
\end{equation}
where all eigenvalues of $A_-\in\R^{m\times m}$ have negative real
part and  $g_-$ and $g_0$ are of second order in $y$ and $z$.  In these
coordinates, we write $b = \bigpar{\begin{smallmatrix} b_- \\ b_0
\end{smallmatrix}}$. 

\begin{myhyp}{(H3)}{Controllability}
$b_0 \neq 0$. 
\end{myhyp}

Since $A_0$ is rotation invariant, we may assume that $b_0 =
\bigpar{\begin{smallmatrix} 0 \\ 1 \end{smallmatrix}}$. 

The next hypothesis is more technical and its meaning will become clear in
Section \ref{sec_bif}. It is, however, generically satisfied.

\begin{myhyp}{(H4)}{Nondegeneracy}
Define the matrix $T=-(A_-^{-1}b_-,A_-^{-2}b_-)\in\R^{m\times 2}$. Let 
$g_\x$ denote the first component of $g_0$, and $h(z)=g_\x(Tz,z)$. 
Then either $\sdpar{h}{\x\x}(0)\neq 0$ or $\sdpar{h}{\x\x\x}(0)<0$. 
\end{myhyp}

Finally, we require the following property of the equilibrium branch
$\fix{x}(\lambda)$:

\begin{myhyp}{(H5)}{Velocity of equilibrium}
The $\y$-component of
$\sdtot{\fix{x}(\lambda)}{\lambda}\evalat{\lambda=0}$ is different from
zero.
\end{myhyp} 

The main result of this paper is the following.

\begin{theorem}
\label{thm_pro1}
Assume that the hypotheses (H1)--(H5) hold. There exist
\begin{itemiz}
\item	strictly positive constants $T$, $M$ and $\kappa$,
\item	a \nbh\ $\cN\subset\cD$ of the origin in $\R^n$,\
\item	a smooth feedback control $u(x,\lambda): \cN\times\ccint{-T}{T} \to
	\R$  with $u(\fix{x}(\lambda),\lambda) = 0$,
\item	and a curve $x_+(\lambda): \ccint{0}{T} \to \R^n$ with
	$\lim_{\lambda\to 0+}
	\norm{x_+(\lambda)-\fix{x}(\lambda)}/\sqrt{\lambda} = K \neq 0$
\end{itemiz}
with the following property. For every $\tau_0\in\coint{-T}{0}$, there
exist strictly positive constants $c_1$ and $\eps_0$ such that, if\/
$0<\eps<\eps_0$, any solution of the equation
\begin{equation}
\label{pro3}
\eps\dtot{x}{\tau} = f(x,\tau) + b u(x,\tau)
\end{equation}
with initial condition $x(\tau_0)\in\cN$ exists on the interval
$\ccint{\tau_0}{T}$ and satisfies the following bounds:
\begin{subequations}
\begin{align}
\label{pro4a}
\norm{x(\tau)-\fix{x}(\tau)} &\leqs M \frac{\eps}{\abs{\tau}}, 
& \tau_0+c_1\eps\abs{\ln\eps} \leqs \tau &\leqs -\eps^{2/3},
\\
\label{pro4b}
\norm{x(\tau)-\fix{x}(\tau)} &\leqs M \eps^{1/3}, 
& -\eps^{2/3} \leqs \tau &\leqs \eps^{2/3},
\\
\label{pro4c}
\norm{x(\tau)-x_+(\tau)} &\leqs M \Bigpar{\frac{\eps}{\tau} +
\frac{\eps^{1/2}}{\tau^{1/4}} \e^{-\kappa\tau^2/\eps}}, 
& \eps^{2/3} \leqs \tau &\leqs T.
\end{align}
\end{subequations}
\end{theorem}

\begin{figure}
 \centerline{\psfig{figure=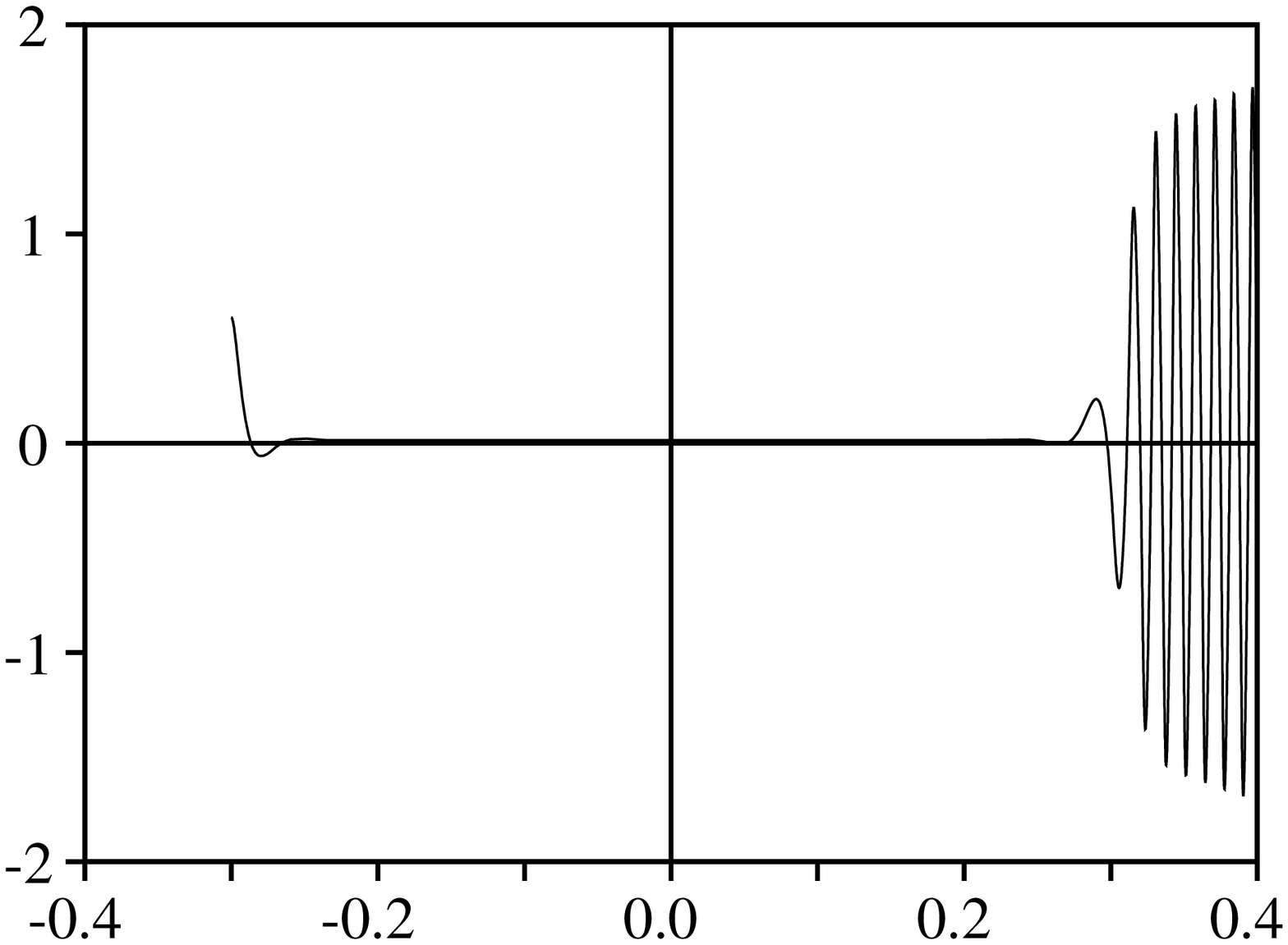,width=68mm,clip=t}
 \psfig{figure=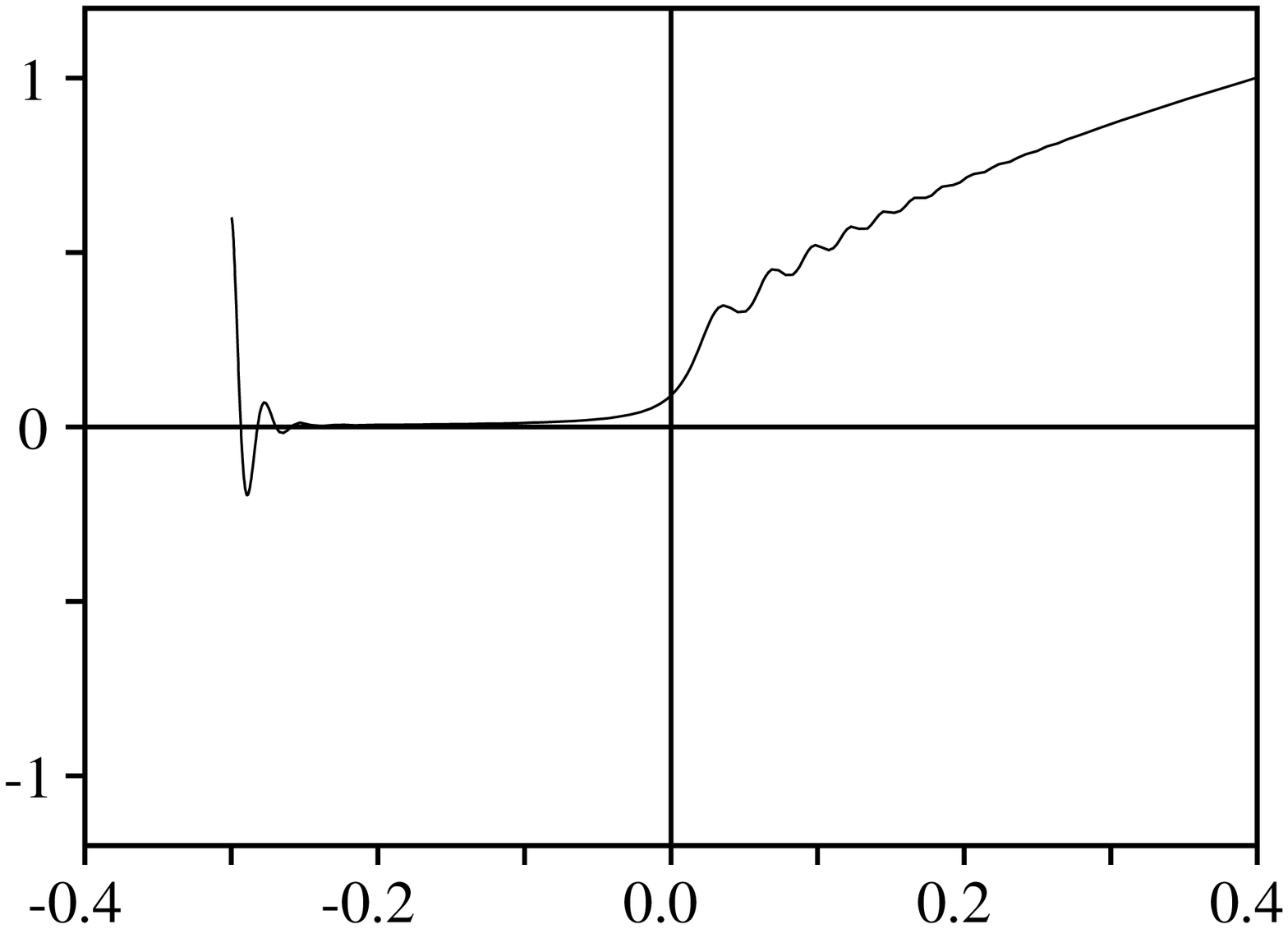,width=68mm,clip=t}}
 \figtext{
 	\writefig	0.3	5.1	a
	\writefig	7.25	5.1	b
 	\writefig	4.15	4.95	$\x$
	\writefig	11.1	4.95	$\x$
	\writefig	13.7	2.8	$\tau$
 }
 \caption[]
 {(a) The slow passage through a Hopf bifurcation (occurring at $\tau=0$)
 leads to the delayed appearance of oscillations. (b) The control we
 construct suppresses this delay, and causes the orbit to track a curve
 lying at a distance of order $\tau^{1/2}$ from the nominal equilibrium 
 (represented by the axis $\x=0$).
 These plots actually show a solution $\x(\tau)$ of equation \eqref{pro6}
 with $a(\tau)=\tau$, $\gamma(\tau)=\delta(\tau)=0$, 
 $R(\x,\y,\tau,\eps)=1$, $\eps=0.003$ and (a) $\mu(\tau)=-0.2$, (b)
 $\mu(\tau) = 2.5\tau$. The eigenvalues of the linearization around the
 origin are $\tau \pm \mu^{1/2}$.}
\label{fig_pro}
\end{figure}

This theorem shows that solutions of \eqref{pro3} will first track the
nominal equilibrium curve $\fix{x}(\tau)$ for negative $\tau$, and then
track a new equilibrium curve $x_+(\tau)$, situated at a distance of order
$\sqrt{\tau}$ from $\fix{x}(\tau)$, for positive $\tau$
(\figref{fig_pro}b). This result looks similar to the stability exchange
for pitchfork bifurcations in \cite{LS}. It is, however, more difficult to
obtain because the specific nature of the bifurcation with double zero
eigenvalue makes the problem intrinsically two-dimensional. In particular,
solutions tend to rotate around the equilibrium branches which are foci
near the bifurcation point. 

The proof is divided into two main steps. In the first one, described in
Section \ref{sec_bif}, we consider the autonomous system \eqref{pro1}. We
construct a feedback $u(x,\lambda)$ in such a way that after some changes
of variables, including a center manifold reduction and a transformation to
normal form, the dynamics is governed by the two-dimensional effective
equation
\begin{equation}
\label{pro5}
\begin{split}
\dtot{\x}{t} &= \y \\
\dtot{\y}{t} &= \mu(\lambda)\x + 2 a(\lambda)\y + \gamma(\lambda)\x^2 +
\delta(\lambda)\x\y - \x^3 - \x^2\y + \Order{4}, 
\end{split}
\end{equation}
where $\mu(0) = \gamma(0) = \delta(0) = 0$ and $\Order{4} =
\Order{\norm{(\x,\y)}^4}$. The function $\mu(\lambda)$ can be controlled to
some extent by the choice of $u$. This so-called cubic Li\'enard equation
happens to be a codimension-4 unfolding of the vector field $(\y,-\x^3
-\x^2\y)$ which has been studied in detail, see \cite{KKR} and references
therein (the authors in \cite{KKR} actually use an equivalent unfolding
obtained by the transformation $\x\mapsto\x+\frac13\gamma$). 

The second step is to show that with this particular feedback, the slowly
time-dependent system \eqref{pro3} can be reduced to 
\begin{equation}
\label{pro6}
\begin{split}
\eps\dtot{\x}{\tau} &= \y  \\
\eps\dtot{\y}{\tau} &= \mu(\tau)\x + 2a(\tau)\y + \gamma(\tau)\x^2 +
\delta(\tau)\x\y - \x^3 - \x^2\y + \Order{4} + \eps R(\x,\y,\tau,\eps).
\end{split}
\end{equation}
The function $R(0,0,\tau,0)$ is related to the drift
$\tdtot{\fix{x}(\tau)}{\tau}$ of the equilibrium branch, and   Hypothesis
(H5) implies that $R(0,0,0,0) \neq 0$.  In Section \ref{sec_sdp} we prove
the following result.

\begin{theorem}
\label{thm_pro2}
Assume that $\mu'(0) > 0$. 
There exist positive constants $d$, $T$, $M$, $\kappa$ and a \nbh\ $\cM$ of
the origin in $\R^2$ with the following property. For every 
$\tau_0\in\coint{-T}{0}$, there is a constant $c_1>0$ such that for
sufficiently small $\eps$, any solution of \eqref{pro6} with initial
condition $(\x,\y)(\tau_0)\in\cM$  satisfies
\begin{align}
\label{pro7}
\abs{\x(\tau)} &\leqs M \frac{\eps}{\abs{\tau}}, & 
\abs{\y(\tau)} &\leqs M \frac{\eps}{\abs{\tau}^{1/2}}, &
&\text{for $\tau_1(\eps) \leqs \tau \leqs 
-\Bigpar{\frac{\eps}{d}}^{2/3}$,} \\ 
\abs{\x(\tau)} &\leqs M \eps^{1/3}, & 
\abs{\y(\tau)} &\leqs M \eps^{2/3}, &
&\text{for $-\Bigpar{\frac{\eps}{d}}^{2/3} \leqs \tau \leqs 
\Bigpar{\frac{\eps}{d}}^{2/3}$},  
\end{align}
where $\tau_1(\eps)=\tau_0+c_1\eps\abs{\ln\eps}$.
If, moreover, the relations 
\begin{equation}
\label{pro9}
\frac{a'(0)}{\mu'(0)} < \frac12, 
\qquad
R(0,0,0,0) \neq 0
\end{equation}
hold, then for $(\eps/d)^{2/3} \leqs \tau \leqs T$ we have
\begin{equation}
\label{pro10}
\begin{split}
\abs{\x(\tau)-\x_+(\tau)} &\leqs M \Bigbrak{\frac{\eps}{\tau} +
\frac{\eps^{1/2}}{\tau^{1/4}}\e^{-\kappa\tau^2/\eps}}, \\ 
\abs{\y(\tau)} &\leqs M \Bigbrak{\frac{\eps}{\tau^{1/2}} +
\eps^{1/2}\tau^{1/4}\e^{-\kappa\tau^2/\eps}}, 
\end{split}
\end{equation}
where 
\begin{equation}
\label{pro11}
\x_+(\tau) = 
\begin{cases}
\sqrt{\mu} + \Order{\tau}, & 
\text{if $R(0,0,0,0) > 0$,} \\
-\sqrt{\mu} + \Order{\tau}, & 
\text{if $R(0,0,0,0) < 0$}
\end{cases}
\end{equation} 
are equilibria of \eqref{pro5}, \ie the right-hand side of \eqref{pro5}
vanishes when $\x=\x_+$ and $\y=0$. 
\end{theorem}


\section{The Uncontrolled Problem}
\label{sec_ucp}

We state a version of Neishtadt's result on bifurcation delay for the
uncontrolled problem
\begin{equation}
\label{ucp1}
\eps\dtot{x}{\tau} = f(x,\tau).
\end{equation}
The results in \cite{Ne,Ne2} apply in fact to a more general slow--fast
system. We are interested in the computation of the delay time. To this end,
we define the function
\begin{equation}
\label{ucp2}
\Psi(\tau) = \int_0^{\tau} \rho(s) \dx s, \qquad
\rho(s) = a(s)+\icx \w(s).
\end{equation}
In some interval $\ooint{-T}{0}$ in which $a(\tau_0)<0$ we can define the
map 
\begin{equation}
\label{ucp3}
\Pi(\tau_0) = \sup_{\tau>\tau_0} \bigsetsuch{\tau}{\re\psi(s) <
\re\psi(\tau_0), \tau_0 < s < \tau},
\end{equation}
giving the first time at which $\re\psi(\tau)$ becomes equal to
$\re\psi(\tau_0)$ again. Observe that $\Pi(\tau_0) > 0$, $\lim_{\tau_0\to 0}
\Pi(\tau)=0$ and $\lim_{\tau_0\to 0}\Pi'(\tau_0)=-1$.

\begin{theorem}
\label{thm_ucp}
Assume that $f(x,\lambda)$ satisfies Hypotheses (H1) and (H2) for all 
$(x,\lambda)$ in some complex \nbh\ of $\cD\times I$. Let $\tau_0<0$ be
such that $a(\tau)<0$ for $\tau_0\leqs\tau<0$. There exist a \nbh\
$\cN$ of $\fix{x}(\tau_0)$, a constant $M>0$, a buffer time $\tau_+>0$
and a continuous function $\delta(\eps)$ going to $0$ as $\eps\to 0$, such
that any solution of \eqref{ucp1} with initial condition $x(\tau_0)\in\cN$
satisfies
\begin{equation}
\label{ucp4}
\norm{x(\tau)-\fix{x}(\tau)} \leqs M\eps \qquad
\text{for $\tau_0 + \delta(\eps) \leqs \tau \leqs \hat{\tau} -
\delta(\eps)$,}
\end{equation}
where the {\em bifurcation delay time} $\hat{\tau}$ is given by 
\begin{equation}
\label{ucp5}
\hat{\tau} = \min\bigset{\Pi(\tau_0), \tau_+}. 
\end{equation}
\end{theorem}

The quantity $\hat{\tau}$ gives in fact a lower bound on the bifurcation
delay. Under more strict assumptions, it also gives an upper bound in the
limit $\eps\to 0$, see \cite{Ne2} and the article by F.\  and M.\ Diener in
\cite{Benoit}. 

The important fact for us is that $\hat{\tau}$ is the minimum of two
quantities. The first one, $\Pi(\tau_0)$, depends on the initial condition
and thus cannot be used to influence the delay. We thus have to modify the
buffer time $\tau_+$, which is determined in the following way.  The
function $\Psi(\tau)$ and the solutions of \eqref{ucp1} can be continued to
a complex \nbh\ of $\tau=0$. For sufficiently small $\abs{\tau_0}$, the
real times $\tau_0$ and $\Pi(\tau_0)$ can be connected by a path
$\Gamma(\tau_0)$ (lying in the upper half plane if $\w_0<0$), along which
$\re\Psi(\tau)$ is constant. One defines a \defwd{negative buffer time}
$\tau_-$ as the smallest real time such that $\Gamma(\tau_-)$ exists and
has certain properties stated in \cite{Ne2}. The positive buffer time is
given by 
\begin{equation}
\label{ucp6}
\tau_+ = \sup_{\tau_-<\tau<0} \Pi(\tau).
\end{equation}
The existence of this buffer time is a rather subtle, nonperturbative
effect, which cannot be understood by naive perturbation theory. The proof
uses an integration of \eqref{ucp1} along a path $\Gamma(\tau)$. 

\begin{example}
\label{ex_ucp}
If $a(\tau)=\tau$ and $\w(\tau)=-\w_0\icx$, then the level lines of 
\begin{equation}
\label{ucp7}
\re\Psi(\tau) = \tfrac12 \bigbrak{(\re\tau)^2 - (\im\tau-\w_0)^2 + \w_0^2}
\end{equation}
are hyperbolas centered at $\tau=\icx\w_0$. Thus we have $\Pi(\tau_0) =
-\tau_0$ and the buffer times are $\tau_{\pm} = \pm\w_0$. The delay may be 
suppressed by choosing a control in such a way that $\w_0=0$. 
\end{example}


\section{Design of the Feedback Control}
\label{sec_bif}

We start by analysing \eqref{pro1} for $\lambda=0$. Coordinates are chosen
as in \eqref{pro2}, where we can scale time in such a way that $\w_0=1$. 
Our feedback control will be constructed to satisfy two requirements:
\begin{enum}
\item	The properties of the bifurcation delay time show that we should
move the imaginary part of the eigenvalues $\pm\w(0)$ towards the origin,
in order to shift the buffer time to the bifurcation time.

\item	This will produce a bifurcation with double zero eigenvalue. In
analogy with works on stabilization of bifurcations \cite{Ae,MS}, the
equilibrium should be stable at the bifurcation point, in order to avoid
escaping trajectories. 
\end{enum}

We claim that an appropriate feedback control has the form 
\begin{equation}
\label{bif2}
u(x,0) = (1-\nu)\x + v(z). 
\end{equation}
[We recall that $x = (y,z) \in \R^m\times\R^2$ and $z=(\x,\y)$.]
The parameter $\nu$ will be ultimately set to $0$, we introduce it in order
to keep track of the effect of an imperfect control. Its aim is to shift
the eigenvalues $\pm\icx\sqrt{\nu}$ of the linearization to the origin. The
function $v(z)$ is a nonlinear term which should assure that the origin of
\eqref{pro1} is stable.  


\subsection{Center Manifold Reduction at $\lambda=0$}
\label{ssec_bifcm}

With the feedback control \eqref{bif2}, equation \eqref{pro1} takes the form
\begin{equation}
\label{bifcm1}
\begin{split}
\dot{y} &= A_-y + \hat{g}_-(y,z) + B_- z \\
\dot{z} &= \hat{A}_0 z + \hat{g}_0(y,z),
\end{split}
\end{equation}
where $\hat{g}_- = g_-(x) + b\, v(z)$, $\hat{g}_0 = g_0(x) + b\, v(z)$,
$B_-=(1-\nu)(b_-,0) \in \R^{m\times 2}$ and the marginally stable part of
the linearization is given by the matrix $\hat{A}_0(\nu) =
\bigpar{\begin{smallmatrix} 0 & 1 \\ -\nu & 0 \end{smallmatrix}}$.

\begin{prop}
\label{prop_bif1}
Equation \eqref{bifcm1} admits an invariant manifold on which the dynamics
is governed by the equation
\begin{equation}
\label{bifcm2}
\dot{z} = \hat{A}_0 z + G(z),
\end{equation}
where $G(z) = \Order{\norm{z}^2}$ is given to third order by equation
\eqref{bifcm7} below.
\end{prop} 

\begin{proof}
The existence of the manifold follows quite directly from the Center
Manifold Theorem \cite{Carr}. We would now like to compute $G$. First, we
introduce the matrix $T\in\R^{m\times 2}$ satisfying $A_-T-T\hat{A}_0 =
-B_-$, which exists because $A_-$ and $\hat{A}_0$ have no eigenvalues in
common \cite{Krein,Wasow}. In fact, it is given by 
\begin{equation}
\label{bifcm3}
T = (t_1,t_2), \quad 
t_1 = A_- t_2, \quad
t_2 = (\nu-1)(A_-^2+\nu\one)^{-1}b_-.
\end{equation}
The change of variables $y = y_1 + Tz$ yields
\begin{align}
\nonumber
\dot{y}_1 &= A_-y_1 + \tilde{g}_-(y_1,z), & 
\tilde{g}_-(y_1,z) &= \hat{g}_-(y_1+Tz,z) - T\hat{g}_0(y_1+Tz,z), \\
\dot{z} &= \hat{A}_0z + \tilde{g}_0(y_1,z), & 
\tilde{g}_0(y_1,z) &= \hat{g}_0(y_1+Tz,z).
\label{bifcm4}
\end{align}
This system admits a center manifold locally described by $y_1=h(z)$, where
$h(z)$ satisfies the partial differential equation
\begin{equation}
\label{bifcm5}
A_-h(z) + \tilde{g}_-(h(z),z) = \sdpar{h}{z}(z) \bigbrak{\hat{A}_0z +
\tilde{g}_0(h(z),z)}. 
\end{equation}
For a vector field $F(x): \R^n \to \R^m$, we denote by $F^{(k)}(x)$ the
terms of order $k$ of its Taylor expansion around $0$. 
We know \cite{Carr} that $h(z)=h^{(2)}(z) + \Order{\norm{z}^3}$, where 
\begin{equation}
\label{bifcm6}
\sdpar{h^{(2)}}{z}(z)\hat{A}_0z - A_-h^{(2)}(z) = \tilde{g}^{(2)}_-(0,z).
\end{equation}
The motion on the center manifold is thus given by \eqref{bifcm2}, where  
\begin{equation}
\label{bifcm7}
\begin{split}
G(z) &= \hat{g}_0(h(z)+Tz,z) \\
&= G^{(2)}(z) + G^{(3)}(z) + \Order{\norm{z}^4},\\
G^{(2)}(z) &= \hat{g}_0^{(2)}(Tz,z) \\
G^{(3)}(z) &= \hat{g}_0^{(3)}(Tz,z) +
\sdpar{\hat{g}}{y}^{(2)}_0(Tz,z)h^{(2)}(z).
\end{split}
\end{equation}
To compute $G^{(3)}(z)$, we need to solve equation \eqref{bifcm6} for
$h^{(2)}$. In fact, we will only need to know that  $h^{(2)}(z) =
-A_-^{-1}\tilde{g}^1\x^2+\Order{\y}$, where $\tilde{g}^1$ is the coefficient
of $\x^2$ in $\tilde{g}^{(2)}_-(0,z)$.  
\end{proof}


\subsection{Stability at $\lambda=0$}
\label{ssec_bifnf}

We intend to construct the nonlinear part of the feedback control of the form
\begin{equation}
\label{bifnf1}
v(z) = v_1\x^2 + v_2\x\y + v_3\y^2 + v_4\x^3,
\end{equation}
with appropriate coefficients $v_i$. They will be determined by the
following result of normal form theory:

\begin{lemma}
\label{lem_bif2}
Consider the system
\begin{equation}
\label{bifnf2}
\begin{split}
\dot{\x} &= \y + c_1\x^2 + c_2\x\y + c_3\y^2 + c_4\x^3 +
\Order{\y\norm{z}^2,\norm{z}^4} \\
\dot{\y} &= \hspace{6.45mm} d_1\x^2 + d_2\x\y + d_3\y^2 + d_4\x^3 +
\Order{\y\norm{z}^2,\norm{z}^4}.
\end{split}
\end{equation}
Let $\alpha = c_1 d_3 + c_4$ and $\beta = 2 c_1^2 + d_4$. Then
\begin{itemiz}
\item	If $d_1 = d_2 + 2c_1 = 0$ and $\alpha, \beta < 0$, the origin is
	asymptotically stable.
\item	Conversely, if the origin is stable, then $d_1 = d_2 + 2c_1 = 0$
	and $\alpha, \beta \leqs 0$.
\end{itemiz}
\end{lemma}

\begin{proof}
The normal form of the 3-jet of \eqref{bifnf2} can be written as 
\begin{equation}
\label{bifnf3}
\begin{split}
\dot{\x} &= \y  \\
\dot{\y} &= \gamma\x\y + \delta \x^2 + \alpha\x^2\y + \beta \x^3,
\end{split}
\end{equation}
with $\gamma = d_2+2c_1$ and $\delta=d_1$. The assertion has been proved in
\cite{Takens}, see also \cite{GH}. In fact, the origin is an unstable
Bogdanov--Takens point  if $\gamma\neq 0$ or if $\delta\neq 0$. If
$\gamma=\delta=0$, it is asymptotically stable if $\alpha$ and $\beta$ are
both negative, and unstable if one of them is positive.
\end{proof}

We would like to choose the coefficients of $v(z)$ in such a way that the
vector field \eqref{bifcm2} satisfies Lemma \ref{lem_bif2}. The next lemma
shows that this is generically possible.

\begin{lemma}
\label{lem_bif3}
Assume that the $\x$-component of $g_0(Tz,z)$ has the form $c_1\x^2 +
c_4\x^3 + \Order{\y,\x^4}$. Then, if either $c_1\neq 0$ or $c_4<0$, one can
find a function $v(z)$ of the form \eqref{bifnf1} such that the origin in
\eqref{bifcm2} is asymptotically stable.
\end{lemma} 

\begin{proof}
Let us consider 
\begin{equation}
\label{bifnf4}
G^{(2)}(z) = \hat{g}^{(2)}_0(Tz,z) = 
\begin{pmatrix}
c_1\x^2 + c_2\x\y + c_3\y^2 \\
(d_1+v_1)\x^2 + (d_2+v_2)\x\y + (d_3+v_3)\y^2 
\end{pmatrix}.
\end{equation}
The first two conditions of Lemma \ref{lem_bif2} are satisfied if
we choose $v_1=-d_1$ and $v_2=-(d_2+2c_1)$. To satisfy the third one, we
have to choose $v_3$ in such a way that $\alpha=c_1(d_3+v_3)+c_4<0$, which
is possible under our assumptions. The last condition looks more difficult
to satisfy, but in fact, we have $\beta=v_4+\mbox{constant}$, where the
constant depends only on previously fixed quantities, so that $\beta$ can
always be made negative.
\end{proof}

The requirement on $g_0(Tz,z)$ is nothing but Hypothesis (H4). 
Note that the coefficients $v_3$ and $v_4$ only have to satisfy
inequalities, while $v_1$ and $v_2$ must have a specific value. We will
discuss in Section \ref{sec_qpr} what happens when these coefficients are
not exactly equal to the prescribed value. 


\subsection{Choice of $u(x,\lambda)$}
\label{ssec_cpu}

We consider now equation \eqref{pro1} for general, fixed values of
$\lambda$. If $\fix{x}(\lambda)$ is the equilibrium branch of $f$, an affine
transformation $x = \fix{x}(\lambda) + S(\lambda)
\bigpar{\begin{smallmatrix}y\\z\end{smallmatrix}}$ yields the system
\begin{equation}
\label{cp1}
\begin{split}
\dot{y} &= A_-(\lambda)y + g_-(y,z,\lambda) +
b_-(\lambda)\tilde{u}(y,z,\lambda) \\
\dot{z} &= A_0(\lambda)z + g_0(y,z,\lambda) +
b_0(\lambda)\tilde{u}(y,z,\lambda), 
\end{split}
\end{equation}
where $A_-(\lambda)$ has eigenvalues with negative real part for
sufficiently small $\lambda$, and 
\begin{equation}
\label{cp2}
A_0(\lambda) = \begin{pmatrix}
a(\lambda) & \w(\lambda) \\ -\w(\lambda) & a(\lambda)
\end{pmatrix}.
\end{equation}
In fact, since $A_0$ is rotation invariant, we may assume that 
$b_0(\lambda)\equiv\bigpar{\begin{smallmatrix}0\\1\end{smallmatrix}}$. 
The terms $g_-$ and $g_0$ are nonlinear.

We now choose the feedback control $u(x,\lambda)$ in such a way that 
\begin{equation}
\label{cpu1}
\tilde{u}(y,z,\lambda) = (1+C\,\lambda)
\bigbrak{(1-\nu)\x+v(z)},
\end{equation}
where $C$ is some constant to be determined, and $v(z)$ has been
constructed in the previous section. In this way, the linear part of
$\dot{z}$ becomes
\begin{equation}
\label{cpu2}
\hat{A}_0(\lambda) = 
\begin{pmatrix}
a(\lambda) & \w(\lambda) \\ -\w(\lambda)+(1-\nu)(1+C\,\lambda) & a(\lambda)
\end{pmatrix}.
\end{equation}
This matrix can be further simplified by a shearing transformation: 
\begin{equation}
\label{cpu3}
z\mapsto\begin{pmatrix}\w^{1/2}&0\\0&\w^{-1/2}\end{pmatrix}z
\qquad \sothat \qquad
\hat{A}_0(\lambda) \mapsto B(\lambda) = 
\begin{pmatrix}
a(\lambda) & 1 \\ \mu(\lambda) & a(\lambda)
\end{pmatrix},
\end{equation}
where 
\begin{equation}
\label{cpu4}
\mu(\lambda) = -\w^2 + (1-\nu)(1+C\,\lambda)\w = 
(C - \w'(0))\lambda - \nu + \Order{\lambda^2+\nu^2}.
\end{equation}
In fact, any of the pairs $(\lambda,\nu)$, $(\lambda,\mu)$ or $(a,\mu)$ can
be considered as independent parameters used to produce the bifurcation
with double zero eigenvalue. We will henceforth set $\nu=0$ and consider
$\bigpar{a(\lambda),\mu(\lambda))=(C - \w'(0))\lambda + \Order{\lambda^2}}$
as a path going through the origin of the two-dimensional parameter space. 

The system \eqref{cp1} admits a center manifold described locally by its 
parametric equation $z=h(y,\lambda)$. On this manifold, the dynamics is
governed by the equation
\begin{equation}
\label{cpu5}
\dot{z} = B(\lambda)z + G(z,\lambda).
\end{equation}
It will not be necessary to compute $G(z,\lambda)$. We only need to know
that $G(z,0)$ has been computed in Proposition \ref{prop_bif1} and
satisfies the requirements of Lemma \ref{lem_bif2}.


\subsection{Normal Forms}
\label{ssec_cpn}

The nonlinear term $G(z,\lambda)$ can be simplified by eliminating all terms
which are not resonant at $\lambda=0$. A convenient basis of resonant terms
up to order $3$ is given by $\set{(0,\x^2),(0,\x\y),(0,\x^3),(0,\x^2\y)}$. 

\begin{figure}
 \centerline{\psfig{figure=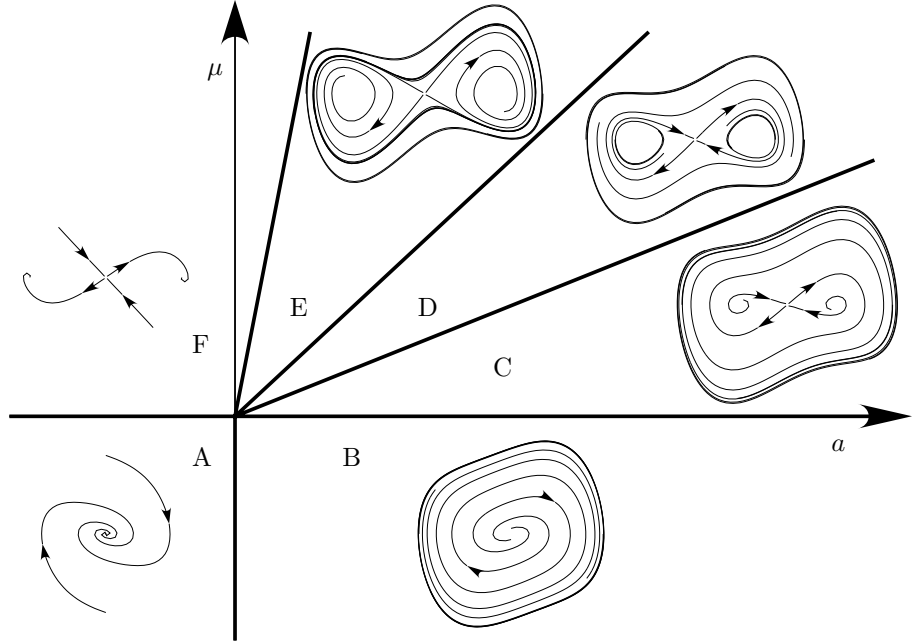,width=120mm,clip=t}}
 \figtext{
 	\writefig	12.3	3.0	$a$
	\writefig	4.0	8.0	$\mu$
	\writefig	3.8	2.8	A
	\writefig	5.8	2.8	B
	\writefig	7.8	4.0	C
	\writefig	6.8	4.8	D
	\writefig	5.1	4.8	E
	\writefig	3.8	4.3	F
 }
 \caption[]
 {Schematic bifurcation diagram of equation \eqref{cp1a} in the plane
 $\hat{\gamma}=\hat{\delta}=0$. The line A-B is the original Hopf
 bifurcation. By moving the eigenvalues' imaginary parts to $0$, we change
 the function $\mu(\lambda)$ in such a way that $\mu(0)=0$. This produces
 new bifurcation lines. The transition A-F is a supercritical saddle-node
 bifurcation, the transition C-B a subcritical one. D-C is a subcritical
 Hopf bifurcation, D-E a homoclinic bifurcation and E-F a saddle-node
 bifurcation of periodic orbits.}
\label{fig_cpb}
\end{figure}

\begin{prop}
\label{prop_cp1}
The $3$-jet of \eqref{cpu5} can be written in either of the following
forms:
\begin{subequations}
\label{cpn}
\begin{equation}
\label{cp1a}
\begin{split}
\dot{\x} &= a\x + \y \\
\dot{\y} &= \mu\x + a\y + \hat{\gamma}\x^2 + \hat{\delta}\x\y - \x^3 -
\x^2\y,
\end{split}
\end{equation}
\begin{equation}
\label{cp1b}
\begin{split}
\dot{\x} &= \y \\
\dot{\y} &= \alpha\x + \beta\y + \gamma\x^2 + \delta\x\y - \x^3 -
\x^2\y,
\end{split}
\end{equation}
\begin{equation}
\label{cp1c}
\begin{split}
\dot{\x} &= \y \\
\dot{\y} &= \mu_1 + \mu_2\x + \mu_3\y + \mu_4\x\y - \x^3 -
\x^2\y.
\end{split}
\end{equation}
\end{subequations} 
Each set of parameters $(a,\mu,\hat{\gamma},\hat{\delta})$,
$(\alpha,\beta,\gamma,\delta)$ or $(\mu_1,\mu_2,\mu_3,\mu_4)$ depends on
$\lambda$ and vanishes at $\lambda=0$.
\end{prop}

\begin{proof}
\eqref{cp1a} is obtained by eliminating nonresonant terms and rescaling
space and time. \eqref{cp1b} comes from the linear transformation
$\y\mapsto\y-a\x$ followed by a scaling. \eqref{cp1c} is obtained with the
transformation $\x\mapsto\x+\frac13\gamma$.
\end{proof}

These normal forms describe equivalent codimension-four unfoldings of the
singular vector field $(\dot{\x}=\y,\dot{\y}=-\x^2\y-\x^3)$. Indeed, the
bifurcation is produced by setting four parameters to zero: the initial
bifurcation parameter $\lambda$, the imaginary eigenvalue shift $\nu$, and
the coefficients $v_1$ and $v_2$ of the feedback control.

Equation \eqref{cp1b} has been studied by Takens in the particular case
$\gamma=\delta=0$ (which occurs when there is a symmetry $z\mapsto -z$), who
obtained the bifurcation diagram shown in \figref{fig_cpb} \cite{Takens,GH}. 
A codimension-three unfolding corresponding to $\mu_4=0$ has been studied
in \cite{VT}. The full codimension-four unfolding \eqref{cp1c} is discussed
in \cite{KKR} (see also references therein). 


\section{Slowly Drifting Parameter}
\label{sec_sdp}

Let us now consider the slowly time-dependent equation \eqref{pro3}. We
write $f(x,\lambda) + b u(x,\lambda) = F(x,\lambda)$, where $u(x, \lambda)$
is the feedback control constructed in the previous section. The equation
thus reads
\begin{equation}
\label{sd1}
\eps\dtot{x}{\tau} = F(x,\tau).
\end{equation}
By construction, $F(x,\tau)$ vanishes on the original equilibrium branch
$\fix{x}(\tau)$. Let us write $F(\fix{x}(\tau)+x_1,\tau) = \hat{A}(\tau)x_1
+ \hat{g}(x_1,\tau)$, with $\hat{g} = \Order{\norm{x_1}^2}$. Hypothesis
(H2) and the properties of $u$ imply that there exists an interval
$\coint{-T_1}{0}$ in which the matrix $\hat{A}(\tau)$ has only eigenvalues
with a negative real part. Moreover, these real parts are bounded away from
0 on any interval $\ccint{\tau_0}{\tau_1}\subset\coint{-T_1}{0}$. It is
well known (see for instance \cite{PR}) that this implies that any
trajectory starting at $\tau_0$ in a sufficiently small \nbh\ of the origin
reaches an $\Order{\eps}$-\nbh\ after a time of order $\eps\abs{\ln\eps}$,
where it remains until $\tau=\tau_1$ (this result is proved using Lyapunov
functions). Hence it suffices to study \eqref{sd1} for $\tau\geqs\tau_1$,
with an initial condition $x(\tau_1) = \Order{\eps}$. 

We first simplify equation \eqref{sd1} by applying similar transformations
as for the time-indepen\-dent system, which will produce some additional
terms of order $\eps$. We will obtain an effective two-dimensional
equation, which we will study by various methods in three different time
intervals. 


\subsection{Reduction of the Equation}
\label{ssec_sdr}

The time-dependent translation $x = \fix{x}(\tau) + x_1$
yields the equation
\begin{equation}
\label{sdr1}
\eps\dtot{x_1}{\tau} = \hat{A}(\tau)x_1 + \hat{g}(x_1,\tau) -
\eps\dtot{\fix{x}(\tau)}{\tau}.
\end{equation}
For small $\tau$, there exists a nonsingular matrix $S(\tau)$ such that 
$S^{-1}\hat{A}S$ is block-diagonal, with blocks $A_-$ and $B$ as in
\eqref{bifcm1} and \eqref{cpu5}. The transformation $x_1 = S(\tau) x_2$
yields the equation
\begin{equation}
\label{sdr2}
\eps\dtot{x_2}{\tau} = 
\begin{pmatrix}
A_-(\tau) & 0 \\ 0 & B(\tau)
\end{pmatrix} x_2 
+ S^{-1}\hat{g}(Sx_2,\tau) - \eps S^{-1}\Bigbrak{\dtot{S(\tau)}{\tau}x_2 +
\dtot{\fix{x}(\tau)}{\tau}}.
\end{equation} 
Next we apply a center manifold reduction. We write $x_2 = (y_2,z_2) \in
\R^m\times\R^2$. 

\begin{prop}
\label{prop_sdr1}
For sufficiently small $\eps$, equation \eqref{sdr2}  admits a local
invariant manifold (local in $z_2$ and $\tau$), with a parametric equation
of the form $x_2 = (h(z_2,\tau,\eps),z_2)$, where $\norm{h(z_2,\tau,\eps)}
\leqs M (\norm{z_2}^2 + \tau^2 + \eps)$. Any solution starting at a
distance of order $\eps$ from this manifold is such that $y_2(\tau) = 
h(z_2(\tau),\tau,\eps) + \Order{\eps}$ in some \nbh\ of
$(x_2,\tau)=(0,0)$, where $z_2(\tau)$ satisfies the equation
\begin{equation}
\label{sdr3} 
\eps\dtot{z_2}{\tau} = B(\tau)z_2 +
G(z_2,\tau) + \eps P(z_2,\tau,\eps), 
\end{equation}
with $B(\tau)$ and $G(z_2,\tau)$ the same functions as in \eqref{cpu5}. 
Moreover, the second component of $P(0,0,0)$ is different from 0.
\end{prop}

\begin{proof}
Equation \eqref{sdr2} can also be written as
\begin{equation}
\begin{split}
\label{sdr4}
\sdtot{y_2}{t} &= A_-(\tau) y_2 + \eps w_-(\tau) + 
\Order{\norm{y_2}^2 + \norm{z_2}^2 + \tau^2 + \eps^2} \\
\sdtot{z_2}{t} &= B(\tau) z_2 + \eps w_0(\tau) + 
\Order{\norm{y_2}^2 + \norm{z_2}^2 + \tau^2 + \eps^2} \\
\sdtot{\tau}{t} &= \eps \\
\sdtot{\eps}{t} &= 0.
\end{split}
\end{equation}
The transformation $y_2 = y_3 - \eps A_-^{-1}(0)w_-(0)$ yields a
system whose linearization at the point $(y_3,z_2,\tau,\eps)=(0,0,0,0)$ is
the matrix
\begin{equation}
\label{sdr5}
\begin{pmatrix}
A_-(0) & 0 & 0 & 0 \\
0 & B(0) & 0 & w_0(0) \\
0 & 0 & 0 & 1 \\
0 & 0 & 0 & 0
\end{pmatrix}.
\end{equation}
The center manifold theorem implies the existence of a local invariant
manifold  $y_3 = \tilde{h}(z_2,\tau,\eps) = \Order{\norm{z_2}^2 + \tau^2 +
\eps^2}$.  Moreover, it is shown in Lemma 1, p.\ 20 of \cite{Carr}, that in
some \nbh\ of the origin, any solution satisfies a bound of the form
\begin{equation}
\label{sdr6}
\norm{y_3(\tau)-h(z_2(\tau),\tau,\eps)} \leqs M\e^{-\kappa (\tau-\tau_0)/\eps} 
\norm{y_3(\tau_0)-h(z_2(\tau_0),\tau,\eps)},
\end{equation} 
for some positive $M,\kappa$. Since
$\norm{y_3(\tau_0)-h(z_2(\tau_0),\tau,\eps)} = \Order{\eps}$, we have 
\begin{equation}
\label{sdr7}
\eps\dtot{z_2}{\tau} = B(\tau)z_2 + g_0(h(z_2,\tau,\eps) +
\Order{\eps}, z_2, \tau, \eps).
\end{equation}
When $\eps=0$, this equation coincides with the equation on the
instantaneous center manifold. The assertion on $P(0,0,0)$ follows from
Hypothesis (H5).
\end{proof}

As a final reduction step, we may apply a nonlinear transformation putting
the 3-jet of $G(z_2,\tau)$ into one of the canonical forms \eqref{cpn}.
This will produce terms of order $\eps\norm{z_2}^2$, that we may also
absorb into  the remainder $P(z_2,\tau,\eps)$. A redefinition of the
variable $\y$ yields the canonical form \eqref{pro6}. In the sequel, we will
use the equivalent form
\begin{equation}
\label{sdr8}
\begin{split}
\eps\dot{\x} &= a(\tau)\x + \y \\
\eps\dot{\y} &= \mu(\tau)\x + a(\tau)\y + \hat{\gamma}(\tau)\x^2 +
\hat{\delta}(\tau)\x\y - \x^3 - \x^2\y
+ \Order{\norm{z}^4} + \eps R(\x,\y,\tau,\eps),
\end{split}
\end{equation}
where we now use the dots to indicate the derivative with respect to
$\tau$. We can rescale time in such a way that
\begin{equation}
\label{sdr9}
a(\tau) = c\tau + \Order{\tau^2}, \quad
\mu(\tau) = \tau + \Order{\tau^2}, \quad
\hat{\gamma}(\tau) = \Order{\tau}, \quad
\hat{\delta}(\tau) = \Order{\tau}, 
\end{equation}
where $c = \tdtot{a}{\mu}(0) = \dot{a}(0)/(C-\dot{\w}(0))$, see \eqref{cpu4}.

We will study equation \eqref{sdr8} by different methods in three different
regions. They are characterized by a constant $d$ which will be chosen
sufficiently small, but is independent of $\eps$. These regions are:
\begin{enum}
\item	Before the bifurcation: $\tau_1(\eps)\leqs\tau\leqs
	-(\eps/d)^{2/3}$, where $\tau_1(\eps) = \tau_0 +
	\Order{\eps\ln\eps}$ is chosen in such a way that $z(\tau_1) =
	\Order{\eps}$. In this region \eqref{sdr8} can be transformed into a
	one-dimensional complex equation, that we study by suitable normal
	form transformations. The main difficulty is to use the averaging
	effect of fast oscillations around the focus. 
\item	During the bifurcation: $-(\eps/d)^{2/3} \leqs \tau \leqs
	(\eps/d)^{2/3}$. In this inner region, \eqref{sdr8} can be reduced
	to a time-dependent Hamiltonian system by an appropriate scaling. 
\item	After the bifurcation: $(\eps/d)^{2/3} \leqs \tau \leqs T$, where
	$T$ will be chosen sufficiently small. By choosing $c$ in an
	appropriate way, we ensure the existence of an attracting
	equilibrium branch. 
\end{enum}


\subsection{Before the Bifurcation}
\label{ssec_sdb}

In the first region, we use a coordinate transformation which diagonalizes
the linear part approximately. It is given by
\begin{equation}
\label{sdb1}
\z = \frac{1}{\sqrt{2}} \bigbrak{\e^{\icx\pi/4}\abs{\mu}^{1/4}\x +
\e^{-\icx\pi/4}\abs{\mu}^{-1/4}\y},
\end{equation}
and its inverse reads
\begin{equation}
\label{sdb2}
\begin{split}
\x &= \tfrac{1}{\sqrt{2}} 
\bigbrak{\e^{-\icx\pi/4}\z+\e^{\icx\pi/4}\cc{\z}\,}
\abs{\mu}^{-1/4} \\
\y &= \tfrac{1}{\sqrt{2}} 
\bigbrak{\e^{\icx\pi/4}\z+\e^{-\icx\pi/4}\cc{\z}\,}
\abs{\mu}^{1/4}. 
\end{split}
\end{equation}
In these variables, equation \eqref{sdr8} becomes
\begin{equation}
\label{sdb3}
\eps\dot{\z} = \rho(\tau)\z 
+ \eps\varphi(\tau)\cc{\z}
+ \eps w(\tau) + G_0(\z,\cc{\z},\tau,\eps),
\end{equation}
where 
\begin{equation}
\begin{split}
&\rho(\tau) = a(\tau)+\icx\sqrt{-\mu(\tau)}+\eps\psi(\tau), 
\qquad
\psi(\tau) = \Order{\abs{\tau}^{-1/2}},
\\
&\varphi(\tau) = \icx\dot{\mu}(\tau)/4\mu(\tau)
 = \Order{\abs{\tau}^{-1}},
\\
&w(\tau) = \frac{1}{\sqrt{2}} \e^{-\icx\pi/4}\abs{\mu}^{-1/4} R(0,0,\tau,0) 
= \Order{\abs{\tau}^{-1/4}}, 
\\
&G_0(\z,\cc{\z},\tau,\eps) = \Order{\abs{\tau}^{1/4}\abs{\z}^2} +
\Order{\abs{\tau}^{-1}\abs{\z}^3} +
\Order{\abs{\tau}^{-5/4}\abs{\z}^4}.
\label{sdb4}
\end{split}
\end{equation}
The main result of this subsection is the following estimate.

\begin{prop}
\label{prop_sdb}
If $d$ and $\eps$ are sufficiently small, there exists a constant $M_1>0$
such that equation \eqref{sdb3} admits a particular solution $\z_0(\tau)$
satisfying 
\begin{equation}
\label{sdb5}
\biggabs{\z_0(\tau) + \eps\frac{w(\tau)}{a(\tau)+\icx\sqrt{-\mu(\tau)}}} 
\leqs M_1\frac{\eps^2}{\abs{\tau}^{9/4}}
\qquad
\mbox{for $\tau_1\leqs\tau\leqs -(\eps/d)^{2/3}$.}
\end{equation}
Moreover, any solution of \eqref{sdb3} with initial condition $\z(\tau_1) =
\Order{\eps}$ satisfies the bound
\begin{equation}
\label{sdb6}
\bigabs{\z(\tau)-\z_0(\tau)} \leqs M_2 \eps
\e^{-\kappa(\tau^2-\tau_1^2)/2\eps}
\end{equation}
on the same time interval, for some positive constants $M_2, \kappa$.
\end{prop}

\begin{cor}
\label{cor_sdb}
Any solution of \eqref{sdr8} with initial condition of order $\eps$
satisfies  
\begin{equation}
\label{sdb6a}
\x(\tau) = \Order{\eps\abs{\tau}^{-1}}, \qquad 
\y(\tau) = \Order{\eps\abs{\tau}^{-1/2}}
\end{equation}
on the interval $\tau_1\leqs\tau\leqs -(\eps/d)^{2/3}$. In particular, at
$\tau=-(\eps/d)^{2/3}$ we have 
\begin{equation}
\label{sdb6b}
\begin{split}
\x &= d^{2/3} \eps^{1/3} 
\bigbrak{R(0,0,0,0) + \Order{d} + \Order{(\eps/d)^{1/3}}}, \\
\y &= d^{1/3}\eps^{2/3} \bigbrak{\Order{d} + \Order{(\eps/d)^{1/3}}}.
\end{split}
\end{equation} 
\end{cor}

\begin{remark}
\label{rem_sdb}
One can in fact show the existence of a particular solution admitting an
asymptotic series of the form
\begin{equation}
\label{sdb6c}
\z_0(\tau) = \frac{\eps}{\abs{\tau}^{3/4}} \Bigbrak{c_0(\tau) +
c_1(\tau)\frac{\eps}{\abs{\tau}^{3/2}} + 
c_2(\tau)\frac{\eps^2}{\abs{\tau}^{3}} + \dotsb}
\end{equation}
For $\tau = (\eps/d)^{2/3}$, we get an asymptotic series in $d$. 
Proposition \ref{prop_sdb} is only a first step, but suffices for our
purposes. 
\end{remark}

The proof of Proposition \ref{prop_sdb} is based on the following two
lemmas. The first one is a rather trivial, but very useful bound on an
integral we will encounter several times, while the second one gives an ``a
priori'' estimate on $\z_0(\tau)$. 

\begin{lemma}
\label{lem_sdb1}
Assume that we are given
\begin{itemiz}
\item	constants $\tau_1\leqs\tau<0$, $a_0, w_0 > 0$ and $p, q$ such
	that $p+1-q > 0$.
\item	$\Psi(\tau): \coint{\tau_1}{0} \to \C$ differentiable
	such that $\re\dot{\Psi}(\tau)\leqs 0$ and
	$\abs{\dot{\Psi}(\tau)}\geqs\abs{\tau}^p/a_0$. 
\item	$w(\tau): \coint{\tau_1}{0} \to \C$ differentiable with  
	$\abs{w(\tau)}\leqs w_0\abs{\tau}^{q-1}$ and
	$\abs{\dot{w}(\tau)}\leqs w_0\abs{\tau}^{q-2}$. 
\end{itemiz}
Then
\begin{equation}
\label{sdb7}
\Bigabs{\e^{\Psi(\tau)/\eps} \int_{\tau_1}^{\tau} \e^{-\Psi(s)/\eps} w(s)\dx
s} \leqs \frac{\eps}{\abs{\tau}^{p+1-q}}K, \qquad
K = a_0 w_0 \Bigbrak{2 + \tfrac{1}{p+1-q}}.
\end{equation}
\end{lemma}

\begin{proof}
Use integration by parts once.
\end{proof}

\begin{lemma}
\label{lem_sdb2}
Let $c_0 > 0$ be given.  
If $\eps$ and $d$ are small enough, there exists a constant $M_1(c_0)$ such
that any solution of \eqref{sdb3} with initial condition
$\abs{\z(\tau_1)}\leqs c_0\eps$ satisfies 
\begin{equation}
\label{sdb8}
\abs{\z(\tau)} \leqs M_1 \frac{\eps}{\abs{\tau}^{3/4}}
\qquad
\mbox{for $\tau_1\leqs\tau\leqs -(\eps/d)^{2/3}$.}
\end{equation}
\end{lemma}

\begin{proof}
Notice that for small enough $d$, we have $\abs{\rho}\geqs
K_0\abs{\tau}^{1/2}$. 

\begin{itemiz}
\item	{\it Step 1: Simplification of the linear part.}

Consider the initial value problem
\begin{equation}
\label{sdb10}
\eps\dot{s} = \bigbrak{\rho(\tau)-\cc{\rho}(\tau)}s + \eps\varphi(\tau) -
\eps\cc{\varphi}(\tau)s^2, \qquad s(\tau_1)=0.
\end{equation}
Since $\sdtot{\abs{s}^2}{\tau} =
(\varphi\cc{s}-\cc{\varphi}s)(1-\abs{s}^2)$, its solution satisfies
$\abs{s(\tau)}\leqs 1$. (In fact, one can prove that $s = 
\Order{\eps\abs{\tau}^{-3/2}}$). The transformation 
$\z = \z_1 + s(\tau)\cc{\z}_1$ yields
\begin{equation}
\label{sdb11}
\eps\dot{\z}_1 = \rho_1(\tau)\z_1 + \eps w_1(\tau) +
G_1(\z_1,\cc{\z}_1,\tau,\eps),
\end{equation}
where $\rho_1 = \rho + \eps\varphi\cc{s}$ satisfies $\abs{\rho_1}\geqs
K_1\abs{\tau}^{1/2}$, and $G_1$ and $w_1$ satisfy similar bounds as $G_0$
and $w$. 

\item	{\it Step 2: Simplification of the cubic part.}

Let us consider the effect of a change of variables
\begin{equation}
\label{sdb12}
\z_1 = \z_2 + h(\z_2,\cc{\z}_2,\tau), \qquad
h(\z_2,\cc{\z}_2,\tau) = \sum_{n+m=3} h_{nm}(\tau)\z_2^n \cc{\z}_2^m.
\end{equation}
It transforms \eqref{sdb11} into
\begin{multline}
\label{sdb13}
\eps\dot{\z}_2 = \rho_1\z_2 + \eps w_1 - \eps\sdpar{h}{\z}w_1 -
\eps\sdpar{h}{\,\cc{\z}}\cc{w}_1 + G_1^{(2)} + \\
+ \bigbrak{G_1^{(3)} + \rho_1h - \sdpar{h}{\z}\rho_1\z_2 -
\sdpar{h}{\,\cc{\z}}\cc{\rho}_1\cc{\z}_2 - \eps\sdpar{h}{\tau}} 
+ \Order{\norm{z_2}^4},
\end{multline}
where $G_1^{(k)}$ denotes terms of order $k$. In particular,
\begin{equation}
\label{sdb14}
G_1^{(3)}(\z_2,\cc{\z}_2,\tau) = \sum_{n+m=3} g_{nm}(\tau)\z_2^n
\cc{\z}_2^m,
\end{equation}
with $g_{nm}(\tau) = \Order{\abs{\tau}^{-1}}$. 
We see that the term in brackets of \eqref{sdb13} can be eliminated if 
\begin{equation}
\label{sdb15}
\eps\dot{h}_{nm} = \bigbrak{(1-n)\rho_1 - m\cc{\rho}_1} h_{nm} + g_{nm}.
\end{equation}
This is a linear equation which can easily be solved. If $(n,m)\neq(2,1)$,
one can choose the initial condition in   such a way that
$\abs{h_{nm}(\tau)}\leqs K_2\abs{\tau}^{-3/2}$. When $(n,m)=(2,1)$,
however, we cannot obtain such a good bound because the imaginary part of
the term in brackets vanishes. Thus we do not attempt to eliminate this
term. We obtain the equation
\begin{equation}
\label{sdb16}
\eps\dot{\z}_2 = \bigbrak{\rho_1(\tau) - g_{21}(\tau)\abs{\z_2}^2} \z_2 +
\eps w_1(\tau) + G_2(\z_2,\cc{\z}_2,\tau,\eps),
\end{equation}
where 
\begin{equation}
\label{sdb17}
\abs{G_2(\z_2,\cc{\z}_2,\tau,\eps)} \leqs 
M_2\bigbrak{\abs{\tau}^{1/4}\abs{\z_2}^2 + 
\eps\abs{\tau}^{-7/4}\abs{\z_2}^2 + \abs{\tau}^{-5/4}\abs{\z_2}^4}.
\end{equation}
Let us write $\rho_2(\tau,\z_2) = \rho_1(\tau) - g_{21}(\tau)\abs{\z_2}^2$.

\item	{\it Step 3: Proof of the bound \eqref{sdb8}.}

The solution of \eqref{sdb16} has to satisfy
\begin{multline}
\label{sdb18}
\z_2(\tau) = 
\e^{\brak{\Psi_2(\tau) - \Psi_2(\tau_1)}/\eps}\z_2(\tau_0) + \\
\e^{\Psi_2(\tau)/\eps} \int_{\tau_1}^{\tau}
\e^{-\Psi_2(\tau)/\eps} \bigbrak{w_1(\tau) + \frac{1}{\eps} 
G_2(\z_2,\cc{\z}_2,\tau,\eps)} \dx s,
\end{multline}
where $\Psi_2(\tau) = \int_0^\tau \rho_2(s,\z_2(s))\dx s$. 
We define the time 
\begin{equation}
\label{sdb19}
\fix{\tau} = \sup_{\ccint{\tau_1}{-(\eps/d)^{2/3}}} 
\bigsetsuch{\tau}{\mbox{$\abs{\z_2(s)}\leqs \sqrt{\eps}$ for $\tau_1\leqs
s\leqs \tau$}}. 
\end{equation}
By continuity, $\fix{\tau} > \tau_1$. 
Using the bounds on $G_2$ we can show that for
$\tau_1\leqs\tau\leqs\fix{\tau}$,
\begin{equation}
\label{sdb20}
\begin{split}
\abs{\rho_2(\tau,\z_2)} &\geqs K_2\abs{\tau}^{1/2}, \\
\abs{G_2} &\leqs M_3 \bigbrak{\eps\abs{\tau}^{1/4} + 
\eps^2 \abs{\tau}^{-7/4}}, \\
\abs{\sdtot{G_2}{\tau}}  &\leqs M_3 
\bigbrak{\eps\abs{\tau}^{-5/4} + \eps^2\abs{\tau}^{-11/4}},
\end{split}
\end{equation}
where $\sdtot{G_2}{\tau} =  \sdpar{G_2}{\tau} + \sdpar{G_2}{\z_2}\dot{\z_2}
+ \sdpar{G_2}{\,\cc{\z}_2}\dot{\cc{\z}}_2$ is estimated using \eqref{sdb16}. 
Using Lemma \ref{lem_sdb1} and the bounds on $w_1$, we obtain that 
\begin{equation}
\label{sdb21}
\begin{split}
\abs{\z_2(\tau)} &\leqs M_4 \bigbrak{\eps\abs{\tau}^{-3/4} +
\eps^2\abs{\tau}^{-9/4}} \\ 
\sothat \quad 
\abs{\z_2(\fix{\tau})} &\leqs M_4 \eps^{1/2} \bigbrak{d^{1/2} +
d^{3/2}}. 
\end{split}
\end{equation}
Taking $d$ sufficiently small, we have $\abs{\z_2(\fix{\tau})} <
\eps^{1/2}$. If we assume that $\fix{\tau} < -(\eps/d)^{2/3}$, we contradict
the definition of $\fix{\tau}$, which shows that $\fix{\tau} =
-(\eps/d)^{2/3}$. 
\end{itemiz}
Going back to the initial variables and using the bounds
on $s$ and $h$, we obtain the conclusion of the lemma.
\end{proof}

\begin{proof}[{\sc Proof of Proposition \ref{prop_sdb}}]
After subtracting $\eps w(\tau)/\rho(\tau)$ from \eqref{sdb3} and
eliminating the term linear in $\cc{\z}$ in the same way as in Lemma
\ref{lem_sdb2}, we obtain 
\begin{equation}
\label{sdb22}
\eps\dot{\z}_1 = \rho_1(\tau)\z_1 + \eps^2w_1(\tau) 
+ G_1(\z_1,\cc{\z}_1,\tau,\eps), 
\end{equation}
where
\begin{equation}
\label{sdb23}
\begin{split}
\abs{\rho_1(\tau)} &\geqs K_1\abs{\tau}^{1/2},\\
w_1(\tau) &= \Order{\abs{\tau}^{-7/4}},\\
G_1(\z_1,\cc{\z}_1,\tau,\eps) &= 
\Order{\abs{\tau}^{1/4}\abs{\z_1}^2} +
\Order{\eps\abs{\tau}^{-7/4}\abs{\z_1}^2} + 
\Order{\abs{\tau}^{-1}\abs{\z_1}^3}.
\end{split}
\end{equation}
Let $\Psi_1(\tau)$ be a primitive of $\rho_1(\tau)$. 
The solution of \eqref{sdb22} with initial condition $\z_1(\tau_1)=0$ 
should satisfy
\begin{equation}
\label{sdb24}
\z_1(\tau) = 
\e^{\Psi_1(\tau)/\eps} \int_{\tau_1}^{\tau}
\e^{-\Psi_1(s)/\eps} \bigbrak{\eps w_1(s) +
\frac{1}{\eps} G_1(\z_1,\cc{\z}_1,s,\eps)} \dx s.
\end{equation}
Using the a priori estimate of Lemma \ref{lem_sdb2} and the bounds
\eqref{sdb23}, we can apply Lemma \ref{lem_sdb1} to estimate the
integral and obtain \eqref{sdb5}. To obtain \eqref{sdb6}, it is sufficient
to subtract the particular solution $\z_0(\tau)$ from the general
solution, and to use the modulus as a Lyapunov function. 
\end{proof}


\subsection{During the Bifurcation}
\label{ssec_sdd}

In this subsection, we study \eqref{sdr8} on the time interval
$\ccint{-(\eps/d)^{2/3}}{(\eps/d)^{2/3}}$. We recall that $d$ is a
constant which will be chosen small, but is independent of $\eps$, while
$\eps$ is small with respect to $d$. In other words, we consider the system
on a time scale $d^{-1}$, which is intermediate between $1$ and $\eps^{-1}$.
In fact it turns out to be useful to take $\eps\leqs d^4$. 

From Corollary \ref{cor_sdb}, we know that at the time $\tau =
-(\eps/d)^{2/3}$, $\x$ is of order $\eps^{1/3}$ and $\y$ is of order
$\eps^{2/3}$. The basic idea to analyse the motion during the bifurcation
is to introduce the scaling of variables
\begin{equation}
\label{sdd1}
\x = (\eps/d)^{1/3} x, \qquad
\y = (\eps/d)^{2/3} y, \qquad
\tau = (\eps/d)^{2/3} t.
\end{equation} 
This scaling transforms the system \eqref{sdr8} into
\begin{equation}
\begin{split}
d\, \dot{x} &= y + (\eps/d)^{1/3} Q_1(x,y,t,\eps), \\
d\, \dot{y} &= t x - x^3 + d R_0 + (\eps/d)^{1/3} Q_2(x,y,t,\eps),
\label{sdd2}
\end{split}
\end{equation}
where $R_0 = R(0,0,0,0)$ and the functions $Q_1$ and $Q_2$ are uniformly
bounded by constants independent of $\eps$ and $d$. This system should be
studied on the time interval $t\in\ccint{-1}{1}$, and with an initial
condition
\begin{equation}
\label{sdd2a}
\begin{split}
x(-1) &= d\, \brak{R_0 + \Order{d} + \Order{(\eps/d)^{1/3}}}
= d R_0 + \Order{d^2},\\
y(-1) &= d\, \brak{\Order{d} + \Order{(\eps/d)^{1/3}}} = \Order{d^2}.
\end{split}
\end{equation}
Equation \eqref{sdd2} is a small perturbation of the Hamiltonian system 
\begin{equation}
\label{sdd3}
H(x,y,t) = \frac{1}{d} \Bigbrak{\frac12 y^2 + \frac14 x^4 - \frac12 t
x^2 - d R_0 x}.
\end{equation}

\begin{lemma}
\label{lem_sdd1}
For $-1\leqs t\leqs 1$, the solution of equation \eqref{sdd2} and of the
Hamiltonian system \eqref{sdd3} with the same initial condition differ by a
term of order $(\eps/d)^{1/3}$. 
\end{lemma}

\begin{proof}
One can use a standard averaging result, see for instance \cite{GH}, Theorem
4.1.1.\ page 168. 
\end{proof}

\begin{figure}
 \centerline{\psfig{figure=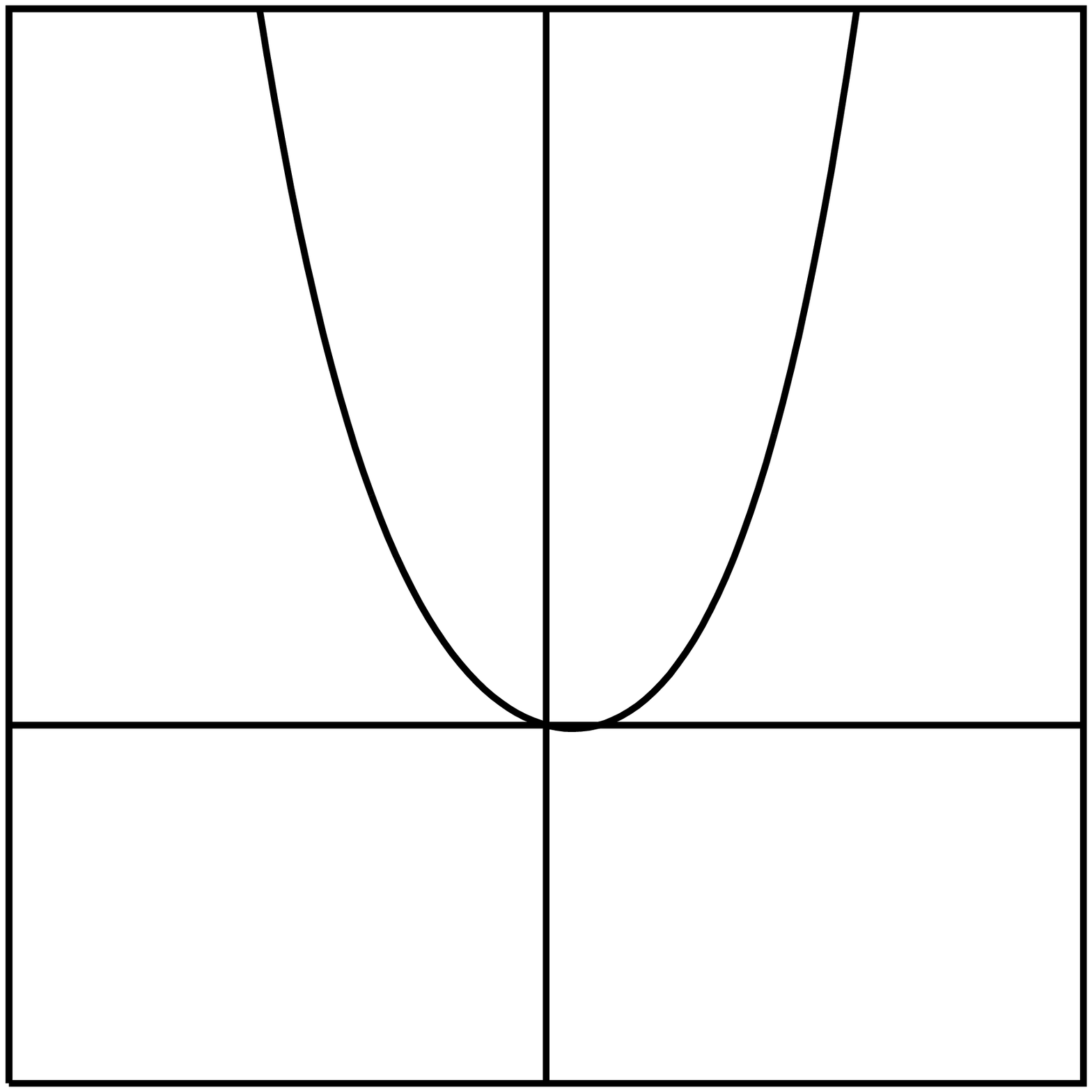,width=40mm,clip=t}
 \hspace{5mm}
 \psfig{figure=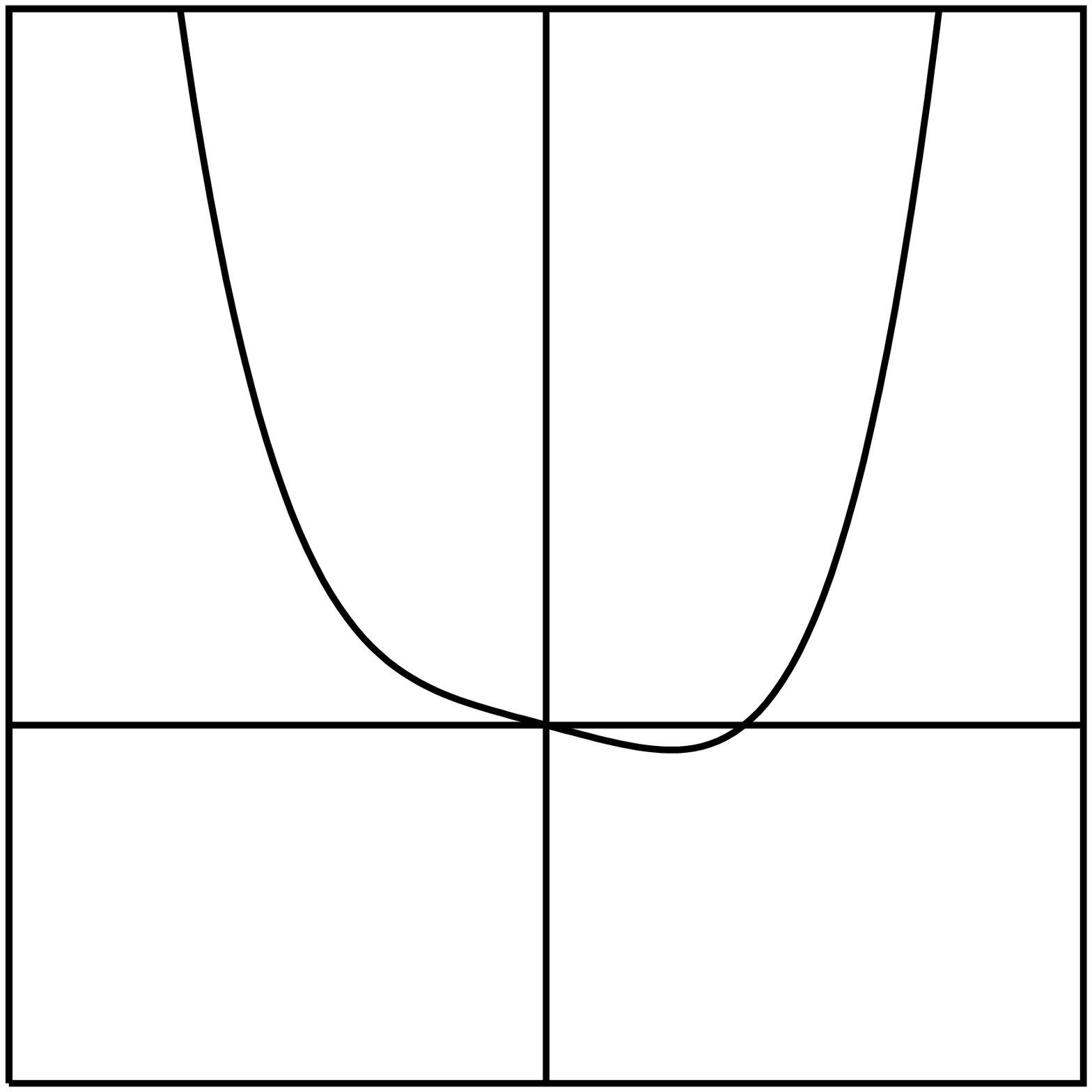,width=40mm,clip=t}
 \hspace{5mm}
 \psfig{figure=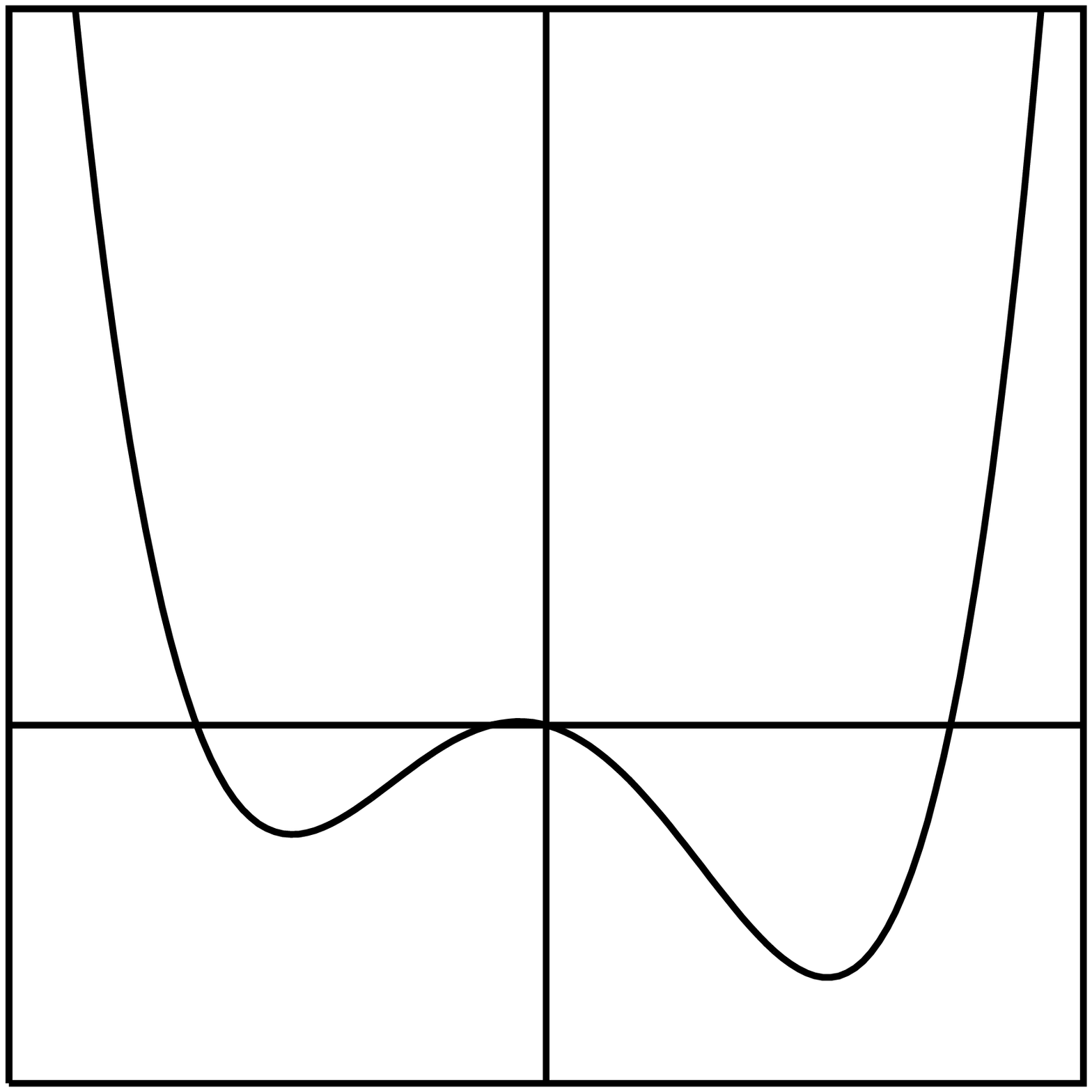,width=40mm,clip=t}}
 \figtext{
 	\writefig	0.3	4.2	a
	\writefig	5.05	4.2	b
	\writefig	9.8	4.2	c
 	\writefig	2.75	4.1	$V$
	\writefig	7.5	4.1	$V$
	\writefig	12.25	4.1	$V$
 	\writefig	4.3	1.5	$x$
	\writefig	9.05	1.5	$x$
	\writefig	13.8	1.5	$x$
 }
 \caption[]
 {The dynamics near the bifurcation point can be reduced to the motion of a
 particle in a time-dependent potential which transforms from a single to a
 double well. The potential $V(x,t) = \frac14 x^4 - \frac12 t x^2 - d R_0
 x$ is shown for $d R_0 = 0.1$ and $t=-1$, $0$ and $1$. Since the energy is
 decreasing, and the barrier between the wells always has a positive
 energy, the particle remains in the same well.}
\label{fig_sdd}
\end{figure}

The Hamiltonian \eqref{sdd3} describes the motion of a particle in a
potential $V(x,t) = \frac14 x^4 - \frac12 t x^2 - d R_0 x$ which changes
from a single well to a double well as $t$ grows from $-1$ to $1$
(\figref{fig_sdd}).  Note that \eqref{sdd2a} implies that the energy is
negative at $t=-1$ if $d$ is sufficiently small. Since 
\begin{equation}
\label{sdd4}
\dtot{H}{t}(x,y,t) = \dpar{H}{t}(x,y,t) = -\frac{1}{2d} x(t)^2,
\end{equation}
the energy must decrease. Since the barrier between the potential wells
always has positive energy, the particle has to remain in the same well,
which is the right one if $R_0 > 0$ and the left one if $R_0 < 0$.

We will however need a more precise estimate, showing that the orbit
remains close to the bottom of the well. This is the main result of this
section. 

\begin{prop}
\label{prop_sdd}
Assume that $R_0\neq 0$. 
For sufficiently small $d$, the solution of the Hamiltonian system
\eqref{sdd3} with the initial condition \eqref{sdd2a} is such that 
\begin{equation}
\label{sdd5}
x(1) = \sign R_0 + \Order{d}, \qquad 
y(1) = \Order{d}.
\end{equation}
\end{prop}
Together with Lemma \ref{lem_sdd1}, this implies

\begin{cor}
\label{cor_sdd}
During the time interval $\ccint{-(\eps/d)^{2/3}}{(\eps/d)^{2/3}}$, the
solution of \eqref{sdr8} with initial condition given by Corollary
\ref{cor_sdb} satisfies $\x = \Order{\eps^{1/3}}$ and $\y =
\Order{\eps^{2/3}}$. At the time $\tau = (\eps/d)^{2/3}$ we have
\begin{equation}
\label{sdd6}
\begin{split}
\x &= (\eps/d)^{1/3} \sign R_0 
+ \Order{\eps^{1/3}d^{2/3}} + \Order{(\eps/d)^{2/3}}, 
\\ 
\y &= \Order{\eps^{2/3}d^{1/3}} + \Order{\eps/d}.
\end{split}
\end{equation}
\end{cor}

\begin{proof}[{\sc Proof of Proposition \ref{prop_sdd}}]
To simplify the notation, we consider the case $R_0 = 1$. 

\begin{itemiz}
\item	{\it Step 1: Transformation of the equation.}
 
Let $\xstar(t)$ be the positive solution of $t\xstar - \xstar^3 + d = 0$.
Then we have 
\begin{gather}
\label{sdd7}
\dxstar(t) = \frac{\xstar(t)}{a(t)}, \qquad
\ddxstar(t) = \frac{2t\xstar(t)}{a(t)^2},
\\
a(t) = 3\xstar^2 - t \geqs\max\set{-t,2d^{2/3},2t}.
\end{gather}
We decrease the order of the drift term\/ $d$ by the transformation
\begin{equation}
\label{sdd8}
\begin{split}
x &= x_1 + \xstar(t) - d^2 \frac{\ddxstar}{a} \\
y &= y_1 + d\, \dxstar(t) - d^3 \dtot{}{t} \Bigpar{\frac{\ddxstar}{a}},
\end{split}
\end{equation}
which implies that $x_1(-1)$, $y_1(-1) = \Order{d^2}$ and yields the system
\begin{equation}
\label{sdd9}
\begin{split}
d\, \dot{x}_1 &= y_1 \\
d\, \dot{y}_1 &= -a_1(t)x_1 - b_1(t)x_1^2 - x_1^3 + d^4 c_1(t),
\end{split}
\end{equation}
where
\begin{equation}
\label{sdd10}
\begin{split}
a_1(t) &= a(t) + \Order{d^{4/3}}, \\
b_1(t) &= 3\xstar(t) + \Order{d}, \\
c_1(t) &= \dtot{^2}{t^2} \Bigpar{\frac{\ddxstar}{a}} + \Order{d^{-5/3}}.
\end{split}
\end{equation}
There is a time $t_\star = \Order{d^{2/3}}$ such that $\dot{a}_1(t)<0$ for
$t<t_\star$ and $\dot{a}_1(t)>0$ for $t>t_\star$.
Using \eqref{sdd7} and some algebra, we obtain the existence of a constant
$M>0$ such that 
\begin{equation}
\label{sdd10a}
\frac{\abs{b_1(t)}}{M} \leqs
\begin{cases}
d \abs{t}^{-1} & \\
d^{2/3} & \\
t^{1/2}, & 
\end{cases}
\quad 
\frac{\abs{c_1(t)}}{M} \leqs
\begin{cases}
d \abs{t}^{-4} \quad & \\
d^{-5/3} & \\
t^{-5/2}, & 
\end{cases}
\quad 
\mbox{for}
\begin{cases}
t\leqs-d^{2/3} &\\
\abs{t}\leqs d^{2/3} &\\
t\geqs d^{2/3}.&
\end{cases}
\end{equation}

\item	{\it Step 2: $t\leqs t_\star$.}

Consider the Lyapunov function
\begin{equation}
\label{sdd11}
V = \Bigbrak{\frac12 y_1^2 + \frac12 a_1(t) x_1^2 + \frac14 x_1^4}^{1/2}.
\end{equation}
We have $V(-1) \leqs M_0 d^2$ and 
\begin{equation}
\label{sdd12}
d\, \dot{V} = 
\frac{1}{2V} 
\Bigbrak{\frac12 d \dot{a}_1 x_1^2 - b_1 x_1^2 y_1 + d^4 c_1 y_1} 
\leqs 2^{-1/2} d^4 \abs{c_1} + 2^{1/2} \frac{b_1}{a_1} V^2.
\end{equation}
From \eqref{sdd10a} we get the bound
\begin{equation}
\label{sdd13}
\int_{-1}^{t_\star} 2^{-1/2} d^3 \abs{c_1(s)} \dx s \leqs M_1 d^2.
\end{equation}
Let $M_2 = M_0 + 2 M_1$ and define the time
\begin{equation}
\label{sdd14}
\hat{t} = \sup_{-1\leqs t\leqs t_\star} 
\bigsetsuch{t}{\mbox{$V(s)<M_2 d^2$ for $-1\leqs s<t$}}.
\end{equation}
For $t\leqs\hat{t}$ we get from \eqref{sdd12} and a standard result on
differential inequalities (see for instance \cite{Hale}) that
\begin{equation}
\label{sdd15}
V(t) \leqs (M_0+M_1) d^2 + M_3 M_2^2 d^{10/3} < M_2 d^2
\end{equation}
for sufficiently small $d$,
which proves that $\hat{t} = t_\star$ and thus $V(t_\star) = \Order{d^2}$. 

\item	{\it Step 3: $t\geqs t_\star$.}

Using the fact that $\dot{a}_1>0$ we obtain
\begin{equation}
\label{sdd16}
d\, \dot{V} \leqs \frac12 d \frac{\dot{a}_1}{a_1} V + 
2^{-1/2} d^4 \abs{c_1} + 2^{1/2} \frac{b_1}{a_1} V^2.
\end{equation}
We choose $M_4$ such that $M_2 d^{1/3} < M_4$ and define 
\begin{equation}
\label{sdd17}
\tilde{t} = \sup_{t_\star\leqs t\leqs 1} 
\bigsetsuch{t}{\mbox{$V(s)<M_4 d^{5/3}$ for $t_\star\leqs s<t$}}.
\end{equation}
For $t\leqs\tilde{t}$, we have 
\begin{equation}
\label{sdd18}
\dot{V} \leqs \frac{\dot{a}_1}{2a_1} \bigbrak{1+M M_4 d^{2/3}} V + 2^{-1/2}
d^3 \abs{c_1}. 
\end{equation}
Using Gronwall's Lemma, we obtain that if
$MM_4d^{2/3}\ln[a_1(1)/a_1(t_\star)]\leqs\ln 2$, 
\begin{equation}
\label{sdd19}
V(t)\leqs 2\biggbrak{\frac{a_1(t)}{a_1(t_\star)}}^{1/2} V(t_\star) + 
\bigbrak{2 a_1(t)}^{1/2} \int_{t_\star}^t d^3
\frac{\abs{c_1(s)}}{a_1(s)^{1/2}} \dx s
\leqs M_5 d^{2/3}.
\end{equation}
Taking $M_4>M_5$, we obtain $\tilde{t}=1$ and thus $V(1)=\Order{d^{5/3}}$. 
Transforming back to the original variables, we obtain $x(1) = \xstar(1) +
\Order{d^{5/3}}$ and $y(1) = d \dxstar(1) + \Order{d^{5/3}}$. \qed
\end{itemiz}
\renewcommand{\qed}{}
\end{proof}


\subsection{After the Bifurcation}
\label{ssec_sda}

In this section, we analyse equation \eqref{sdr8} for $\tau\geqs
(\eps/d)^{2/3}$. It is in fact more convenient to use the form 
\begin{equation}
\label{sda1}
\begin{split}
\eps\dot{\x} &= \y \\
\eps\dot{\y} &= \bar{\mu}(\tau)\x + 2a(\tau)\y + \gamma(\tau)\x^2 +
\delta(\tau)\x\y - \x^3 - \x^2\y
+ \Order{\norm{z}^4} + \eps R(\x,\y,\tau,\eps),
\end{split}
\end{equation}
obtained by the transformation $\y\mapsto\y-a(\tau)\x$, where $\bar{\mu} =
\mu-a^2$. The right-hand side vanishes approximately on three curves
$(\x,\y)\equiv(0,0)$ and
$(\x,\y)=(\fix{\x}_{\pm}(\tau),0)$, where 
\begin{equation}
\label{sda2}
\fix{\x}_{\pm}(\tau) = \pm\bar{\mu}^{1/2} + \Order{\tau}
= \pm \tau^{1/2} + \Order{\tau}.
\end{equation} 
We assume here that $R(0,0,0,0) > 0$ so that the initial condition is close
to $\fix{\x}_+$, the other case is obtained by symmetry. The translation $\x =
\fix{\x}_+(\tau) + \x_1$, $\y = \y_1$ yields
\begin{equation}
\label{sda3}
\begin{split}
\eps\dot{\x_1} &= \y_1  \\
\eps\dot{\y_1} &= -\beta \x_1 + 2\alpha \y_1 + \tilde{\gamma} \x_1^2 +
\tilde{\delta} \x_1\y_1 - \x_1^3 - \x_1^2\y_1
+ \Order{\norm{z_1}^4} + \eps R,
\end{split}
\end{equation}
where (see equation \eqref{sdr9})
\begin{equation}
\label{sda4}
\begin{split}
\alpha(\tau) &= a - \tfrac12\bar{\mu} + \tfrac12(\gamma-\delta)\fix{\x}_+ =
(c-\tfrac12)\tau + \Order{\tau^{3/2}}, \\
\beta(\tau) &= 2 {\fix{\x}_+}^2 - \delta \fix{\x}_+ = 2\tau +
\Order{\tau^{3/2}}, \\
\tilde{\gamma}(\tau) &= \gamma - 2 \fix{\x}_+ = -2\tau^{1/2} +
\Order{\tau}, \\
\tilde{\delta}(\tau) &= \delta - 3 \fix{\x}_+ = -3\tau^{1/2} +
\Order{\tau}, \\
w_1(\tau) &= -\sdtot{}{\tau} \fix{\x}_+(\tau) = -\tfrac12 \tau^{-1/2} +
\Order{1}.
\end{split}
\end{equation}
Notice that the linearization of \eqref{sda3} admits the eigenvalues $\alpha
\pm\icx\beta^{1/2}$. It can be diagonalised approximately by the transformation
\begin{equation}
\label{sda5}
\z = \frac{1}{\sqrt2} \bigbrak{\e^{\icx\pi/4}\beta^{1/4} -
\e^{-\icx\pi/4}\alpha\beta^{-1/4}}\x_1 + \frac{1}{\sqrt2}
\e^{-\icx\pi/4}\beta^{-1/4}\y_1.
\end{equation}
We obtain a system similar to \eqref{sdb3}:
\begin{equation}
\label{sda6}
\eps\dot{\z} = \rho(\tau)\z + \eps\varphi(\tau)\cc{\z} + \eps w(\tau) +
g(\z,\cc{\z},\tau,\eps),
\end{equation}
where 
\begin{equation}
\label{sda7}
\begin{split}
\rho(\tau) &= \alpha + \icx\beta^{1/2} +
\tfrac12\icx\eps\dot{\alpha}\beta^{-1/2}, \\
\varphi(\tau) &= \tfrac14\icx\dot{\beta}\beta^{-1} -
\tfrac12\dot{\alpha}{\beta^{-1/2}} 
= \Order{\tau^{-1}}, \\
w(\tau) &= 
\tfrac{1}{\sqrt2} \e^{-\icx\pi/4}\beta^{-1/4} R(0,0,\tau,0) 
+ \Order{\tau^{1/4}}, \\
g(\z,\cc{\z},\tau,\eps) &= \Order{\tau^{-1/4}\abs{\z}^2} +
\Order{\tau^{-1}\abs{\z}^3} + \Order{\tau^{-5/4}\abs{\z}^4}.
\end{split}
\end{equation}
Finally, by Corollary \ref{cor_sdd}, we get the following estimate on the
initial condition:
\begin{equation}
\label{sda8}
\bigabs{\z((\eps/d)^{2/3})} \leqs M_0 d^{1/2}\eps^{1/2}.
\end{equation}
The main result of this subsection is the following.

\begin{prop}
\label{prop_sda}
Assume that $c = \tdtot{a}{\mu}(0) < \frac12$. Then there exist positive
constants $M$, $\kappa$ and $T$ such that, if $\eps$ and $d$ are
sufficiently small, any solution of \eqref{sda6} with initial condition
\eqref{sda8} satisfies the bounds
\begin{equation}
\label{sda9}
\bigabs{\z(\tau)} \leqs M \bigbrak{\eps\tau^{-3/4} +
\eps^{1/2}\e^{-\kappa\tau^2/\eps}} 
\qquad
\mbox{for $(\eps/d)^{1/3} \leqs \tau \leqs T$.}
\end{equation}
\end{prop}

\begin{cor}
\label{cor_sda}
On the same time interval, there is a constant $M_1>0$ such that the
solution of \eqref{sda1} satisfies
\begin{equation}
\label{sda10}
\begin{split}
\bigabs{\x(\tau) -  \fix{\x}_+(\tau)} &\leqs 
M_1 \bigbrak{\eps\tau^{-1} + \eps^{1/2}\tau^{-1/4}\e^{-\kappa\tau^2/\eps}}, 
\\
\bigabs{\y(\tau)} &\leqs 
M_1 \bigbrak{\eps\tau^{-1/2} + \eps^{1/2}\tau^{1/4}\e^{-\kappa\tau^2/\eps}}. 
\end{split}
\end{equation}
\end{cor}
The proof of Proposition \ref{prop_sda} follows directly from the two lemmas
given below.

\begin{lemma}
\label{lem_sda1}
Assume that $c < \frac12$. There are positive constants $M$ and $T$ such
that \eqref{sda6} admits a particular solution $\z_0(\tau)$ satisfying 
\begin{equation}
\label{sda11}
\bigabs{\z_0(\tau)} \leqs M \eps\tau^{-3/4}  
\qquad
\mbox{for $(\eps/d)^{1/3} \leqs \tau \leqs T$.}
\end{equation}
\end{lemma}

\begin{proof}
The proof is similar to the proof of Lemma \ref{lem_sdb2}, so we only
outline the differences.
\begin{enum}
\item	We have to eliminate quadratic terms from the equation as well. In
	order to get a normal form similar to \eqref{sdb16}, we start by
	eliminating quadratic terms, then we remove the term linear in
	$\cc{\z}$, and then only the nonresonant cubic terms.

\item	For $(\eps/d)^{1/3} \leqs \tau \leqs \eps^{1/2}$, we eliminate the
	real part of the linear term by the transformation $\z =
	\exp[c\tau^2/\eps]\z_1$. We then change the direction of time, fix
	a $\z_1(\eps^{1/2})$ of order $\eps^{5/8}$ and use Lemma
	\ref{lem_sdb2}. 
	 
\item	For $\tau\geqs\eps^{1/2}$, we proceed exactly as in Lemma
	\ref{lem_sdb2}.	\qed
\end{enum}
\renewcommand{\qed}{}
\end{proof}

\begin{lemma}
\label{lem_sda2}
Any solution of \eqref{sda6} with initial condition \eqref{sda8} satisfies 
\begin{equation}
\label{sda12}
\bigabs{\z(\tau)-\z_0(\tau)} \leqs M \eps^{1/2}\e^{-\kappa\tau^2/\eps} 
\qquad
\mbox{for $(\eps/d)^{1/3} \leqs \tau \leqs T$.}
\end{equation}
\end{lemma}

\begin{proof}\hfill

\begin{itemiz}
\item	{\it Step 1: Hamiltonian system.}

Consider, as a special case of \eqref{sda3}, the Hamiltonian system
\begin{equation}
\label{sda13}
\begin{split}
\eps\dot{\x}_1 &= \y_1 - \tfrac12 \eps \tau^{-1/2} \\
\eps\dot{\y}_1 &= -2\tau\x_1 - 3\sqrt{\tau}\x_1^2 - \x_1^3.
\end{split}
\end{equation}
Lemma \ref{lem_sda2} shows the existence of a particular solution
$\x_0(\tau) = \Order{\eps\tau^{-1}}$, $\y_0(\tau) =
\Order{\eps\tau^{-1/2}}$. If $(\x_1,\y_1) = (\x_0,\y_0) + (\x_2,\y_2)$, the
dynamics of $(\x_2,\y_2)$ is governed by a Hamiltonian of the form
\begin{equation}
\label{sda14}
H(\x_2,\y_2,\tau) = \frac12 \y_2^2 + \tau k_1(\tau)\x_2^2 +
\sqrt{\tau}k_2(\tau)\x_2^3 +\frac14\x_2^4,
\end{equation}
where $k_1(\tau)$ and $k_2(\tau)$ are bounded functions. Using
\eqref{sda8}, one can show that for sufficiently small $d$, there exists a
constant $M_0$ such that 
\begin{equation}
\label{sda15}
H(\tau) \leqs M_0 d^2 \tau^2.
\end{equation}

\item	{\it Step 2: Normal forms.}

We write \eqref{sda6} in the form
\begin{equation}
\label{sda16}
\eps\dot{\z} = \rho(\tau)\z + \eps\varphi(\tau)\cc{\z} + \eps w(\tau) +
g^0(\z,\cc{\z},\tau) + g^1(\z,\cc{\z},\tau,\eps), 
\end{equation}
where $g^0(\z,\cc{\z},\tau)$ is the contribution of the Hamiltonian
approximation \eqref{sda13} of \eqref{sda3}, and 
\begin{equation}
\label{sda17}
g^1(\z,\cc{\z},\tau,\eps) = \Order{\tau^{1/4}\abs{\z}^2} +
\Order{\tau^{-1/2}\abs{\z}^3} + \Order{\tau^{-5/4}\abs{\z}^4}.
\end{equation}
This relation holds because the full system \eqref{sda3} is a perturbation
of size $1+\sqrt{\tau}$ of the Hamiltonian system \eqref{sda13}. 

We now perform a number of changes of variables: a translation $\z =
\z_0(\tau) + \z_1$, where $\z_0(\tau)$ is the particular solution of Lemma
\ref{lem_sda1}; a linear change of variables $z_1 = \z_2 +
s(\tau)\cc{\z}_2$, where $s$ satisfies \eqref{sdb12}, 
which cancels the term linear in $\cc{\z}_1$; and a transformation to normal
form $\z_2 = \z_3 + h(\z_3,\cc{\z}_3,\tau)$ which yields the equation
\begin{equation}
\label{sda18}
\eps\dot{\z}_3 = \rho_3(\tau)\z_3 + c^0(\tau;\eps) \abs{\z_3}^2 \z_3 +
c^1(\tau;\eps) \abs{\z_3}^2 \z_3 + \Order{\tau^{-5/4}\abs{\z_3}^4},
\end{equation}
where the functions $c^0 = \Order{\tau^{-1}}$ and $c^1 = \Order{\tau^{-1/2}}$ 
denote contributions of $g^0$ and $g^1$, respectively. 

\item	{\it Step 3: Bounds on the coefficients.}

We claim that 
\begin{equation}
\label{sda19}
\re \rho_3(\tau) = \alpha(\tau) + \Order{\eps\tau^{-1/2}}, 
\qquad
\re c^0(\tau) = \Order{\tau^{-1/2}}. 
\end{equation}
The first claim can be checked by a direct calculation. We observe that the
linearization of \eqref{sda3} around the particular solution $(\x_0,\y_0)$
has the form $\bigpar{\begin{smallmatrix}0&1\\
-\tilde{\beta}&2\tilde{\alpha}  \end{smallmatrix}}$, where $\tilde{\alpha}
= \alpha + \Order{\eps\tau^{-1/2}}$ and $\tilde{\beta} = \beta + 
\Order{\eps\tau^{-1/2}}$. Then we show that the function $s(\tau)$ occuring
in the linear transformation is such that $\im s(\tau) =
\Order{\eps\tau^{-1}}$.

The second claim can be proved without lengthy calculations. By
construction, $\tau c^0(\tau)$ is polynomial in
$\set{\tau^{1/2},\eps\tau^{-3/2}}$. If we assume by contradiction that the
leading term of $\re c^0(\tau)$ is of order $\tau^{-1}$, we reach the
conclusion that $\z_2$ would grow faster than allowed by the estimate
\eqref{sda15}. 

\item	{\it Step 4: Final estimate.}

The Lyapunov function $V = \abs{z_3}^2$ satisfies the equation
\begin{equation}
\label{sda20}
\eps\dot{V} \leqs -2\kappa\tau V \bigbrak{1 - M_2\tau^{-3/2}V -
M_2\tau^{-9/4}V^{3/2}},
\end{equation}
where $\kappa = \frac12 - c$ and $M_2>0$. We obtain the conclusion in a
similar way as in Proposition \ref{prop_sdd}. 		\qed
\end{itemiz}
\renewcommand{\qed}{}
\end{proof}


\section{Qualitative properties and robustness}
\label{sec_qpr}

\begin{figure}
 \centerline{\psfig{figure=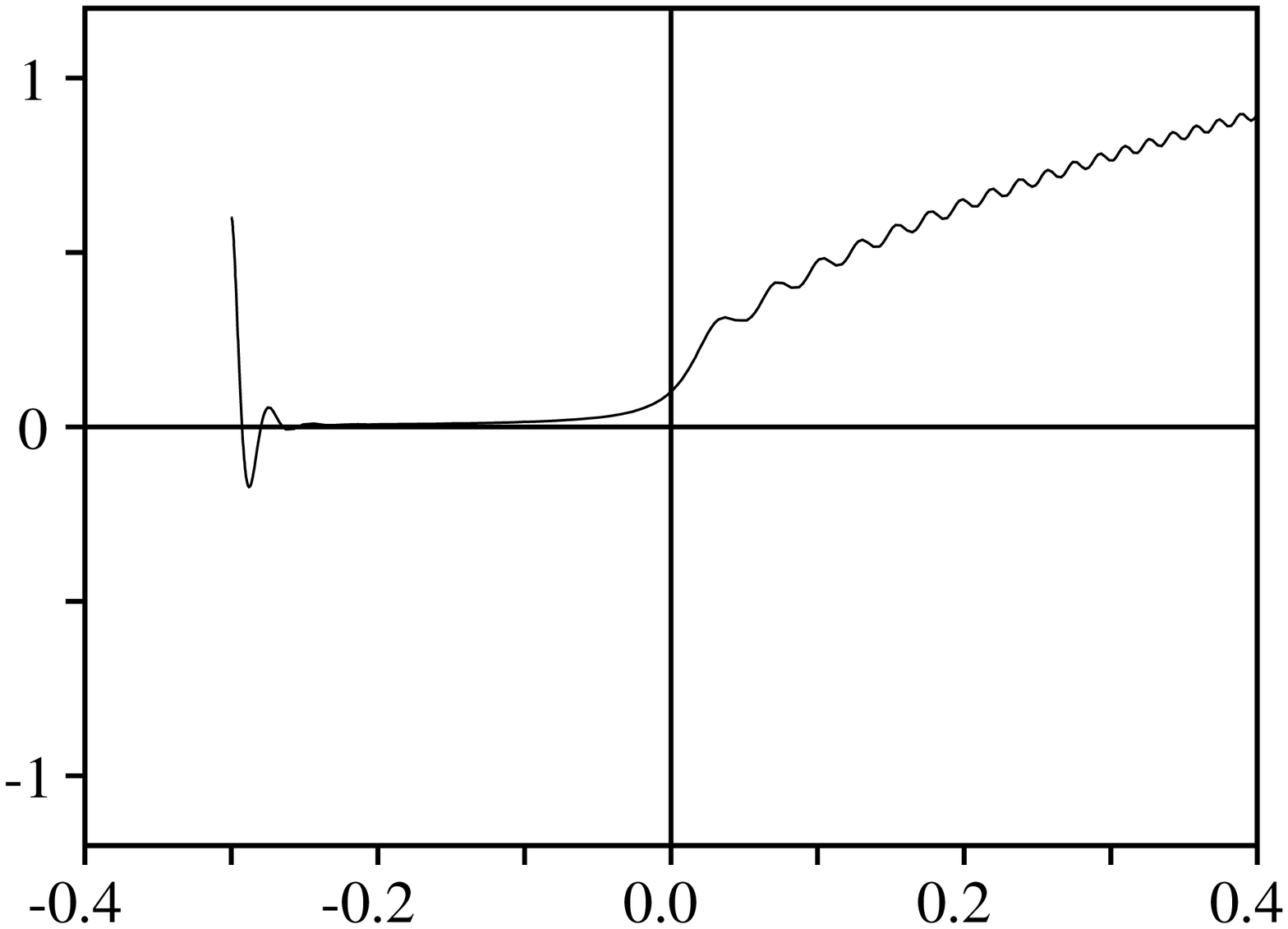,width=68mm,clip=t}
 \psfig{figure=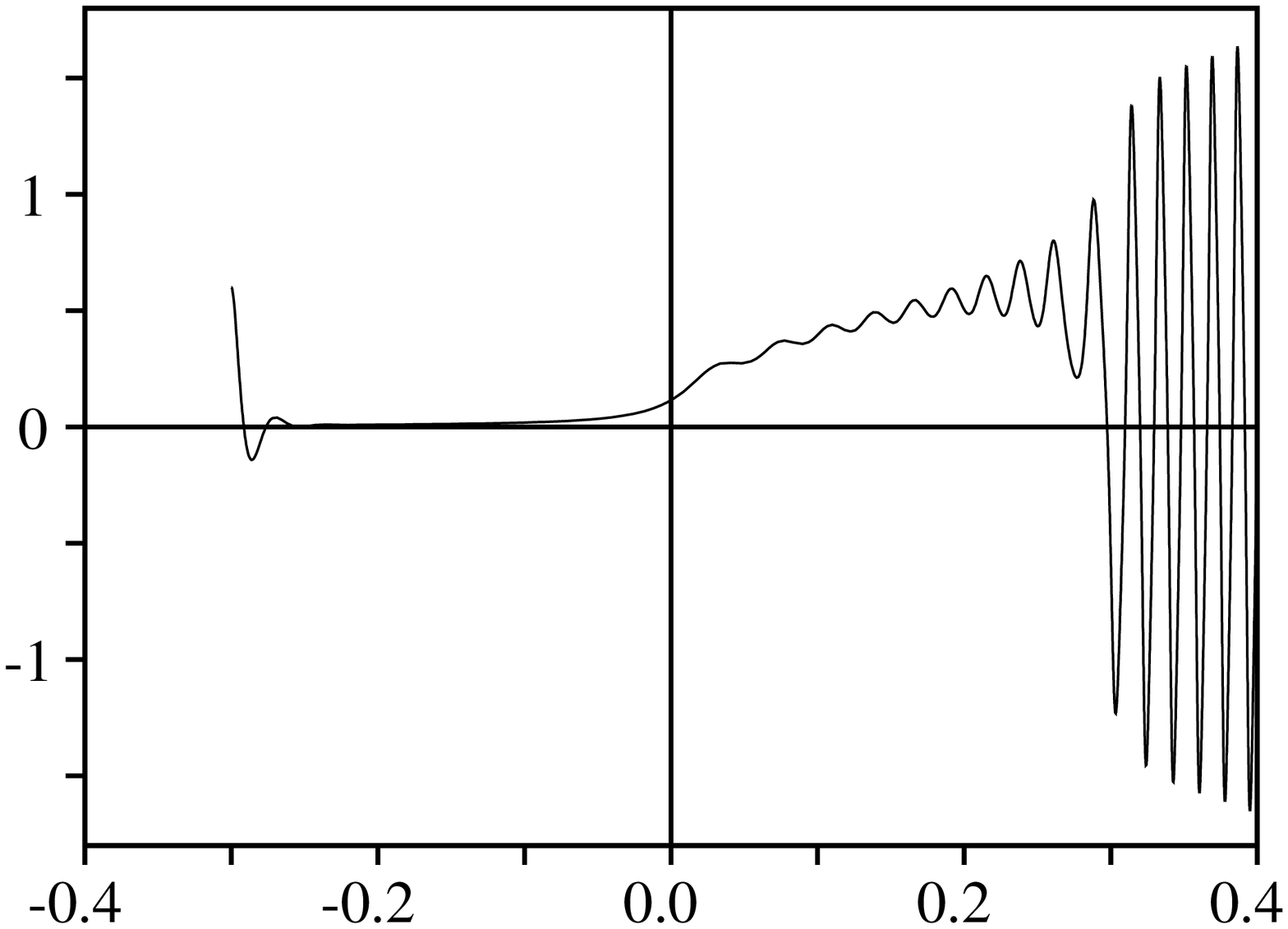,width=68mm,clip=t}}
 \figtext{
 	\writefig	0.3	5.1	a
	\writefig	7.25	5.1	b
 	\writefig	4.15	4.95	$\x$
	\writefig	11.1	4.95	$\x$
	\writefig	6.75	2.8	$\tau$
 }
 \caption[]
 {Same as \figref{fig_pro}, but for values of $\tdtot{a}{\mu}$ to which
 Theorem \ref{thm_pro2} does not apply. (a) When $\mu(\tau)=2\tau$, we move
 just along the C-D line of \figref{fig_cpb}; the equilibria are marginally
 stable, thus the solution keeps oscillating around them with a constant
 amplitude. (b) When $\mu(\tau)=0.5\tau$, we go from region A to region C;
 now the asymmetric equilibria are unstable; the trajectory oscillates for
 some time around them because of Proposition \ref{prop_sdd}, but ultimately
 escapes to the periodic orbit.}
\label{fig_qpr}
\end{figure}

We conclude by discussing a few cases to which Theorem \ref{thm_pro2} does
not apply. 

\begin{itemiz}
\item	{\it Other values of $\tdtot{a}{\mu}$:}

The result requires a sufficiently low value of $\tdtot{a}{\mu}$, which
should be such that we traverse the bifurcation diagram \figref{fig_cpb}
from region A to one of the regions D, E or F. Note that in a \nbh\ of the
bifurcation point, the qualitative behaviour is independent of the functions
$\gamma$ and $\delta$. The only relevant fact is that the bifurcating
equilibrium branches should be attracting. This fact is not surprising
when we steer into regions E or F. It is a bit more surprising when we go
into region D, because the equilibrium is surrounded by an unstable periodic
orbit. The fact that the trajectory lands inside this orbit is due to the
specific nature of the Hamiltonian approximation valid near the bifurcation
point.

If we further increase $\tdtot{a}{\mu}$ so as to reach region C, the
equilibria are no longer attracting. On the C-D boundary, trajectories
oscillate around these equilibria with approximately constant amplitude
(\figref{fig_qpr}a), while in the region C, they depart from them after a
time of order $\eps^{1/2}$ and finally reach the outer periodic orbit
(\figref{fig_qpr}b). In view of our control problem, this behaviour is not
desirable in the sense that we have an appearance of large amplitude
oscillations, although the delay is not macroscopic. 

\item	{\it Imperfect control:}

While constructing the feedback control in Section \ref{sec_bif}, we had to
adjust precisely three parameters: the imaginary eigenvalue shift $\nu$ and
the two parameters $v_1$ and $v_2$ of the nonlinear part. If these
parameters are not set exactly to the desired value, the dynamics will still
be governed by equation \eqref{pro6}, but the functions $a(\tau)$,
$\mu(\tau)$, $\gamma(\tau)$ and $\delta(\tau)$ will not vanish exactly at
the same time. The same is true to some extent if the feedback $u(x,\tau)$
does not vanish exactly on the nominal equilibrium branch $\fix{x}(\tau)$. 
In other words, such imperfections result in the fact that we traverse the
bifurcation diagram of \figref{fig_cpb} on a curve which misses the origin. 
Two situations may occur:

If we traverse the diagram on a path A-F-E, we first experience a pitchfork
bifurcation, which results in an exchange of stabilities as shown in
\cite{LS}. Thus the trajectory will still track a stable equilibrium, but
the bifurcation will occur a little bit earlier.

If we traverse the diagram on a path A-B-C, the situation is less favourable.
The Hopf bifurcation A-B will not be felt immediately, but when the region C
is reached, there is a risk that the trajectory jumps to the periodic orbit.
This behaviour can only be avoided if the slope $\tdtot{\mu}{a}$ is large
enough that the region D is reached before the trajectory has departed from
the asymmetric branch. 
\end{itemiz}


\end{document}